\documentclass[a4paper,12pt]{article}
\usepackage{graphicx,rotating,hyperref,slashed,verbatim,amsmath,xcolor,amssymb,amsfonts,expdlist,colortbl,youngtab}
\usepackage{amsfonts}
\usepackage{dsfont}
\usepackage[utf8]{inputenc}
\usepackage{amsthm}
\usepackage{halloweenmath}
\usepackage{physics}
\usepackage{amsmath}
\usepackage{physics}
\usepackage{mathtools}
\usepackage{hhline}
\usepackage{comment}
\usepackage{colortbl}
\usepackage{multicol}
\usepackage{pdflscape}
\usepackage{multirow}
\usepackage{lscape}
\usepackage{placeins}
\usepackage{booktabs}    
\usepackage{calc}        
\usepackage{ytableau}
\usepackage{tikz}
\usepackage[compat=1.1.0]{tikz-feynman}
\usepackage{subcaption}
\usepackage{fancyhdr}
\usepackage{cleveref}
\usepackage{cancel}

\usepackage[utf8]{inputenc}
\usepackage{csquotes}
\usepackage[backend=biber,style=numeric-comp,sorting=none, maxnames=5]{biblatex}
\newcommand{\Dslash}{\slashed{D}}
\newcommand{\vslash}{\slashed{v}}
\newcommand{\ccdot}{\!\cdot\!}
\newcommand{\vcD}{v \ccdot D}

\newcommand{\amp}[3]{\frac{\langle #1(p') |\, #2\, | #3(p) \rangle}{\sqrt{m_{\let\overline\relax#3} m_{#1}}}}

\hypersetup{colorlinks,bookmarksopen,bookmarksnumbered,
	linkcolor=blus,pdfstartview=FitH,urlcolor=rossos,citecolor=verde}

\usepackage{tabularx}

\definecolor{gold}{RGB}{187,161,79}
\definecolor{silver}{RGB}{192,192,192}
\definecolor{darkgreen}{RGB}{0, 130, 0}

\newcommand{\sfootnote}[1]{}
\definecolor{bluc}{cmyk}{1,1,0,0.1}
\definecolor{rossoCP3}{cmyk}{0,.88,.77,.40}
\definecolor{rosso}{cmyk}{0,1,1,0.4}
\definecolor{giallo}{cmyk}{0,.33,1,0}
\definecolor{rossos}{cmyk}{0,1,1,0.55}
\definecolor{rossoc}{cmyk}{0,1,1,0.2}
\definecolor{verdes}{cmyk}{0.92,0,0.59,0.4}

\hypersetup{colorlinks, bookmarksopen, bookmarksnumbered,
	citecolor=verdes, linkcolor=bluc, pdfstartview=FitH, urlcolor=black!35!blue}

\newcommand{\mio}[1]{}

\definecolor{Gray}{gray}{0.95}

\usepackage{color}
\definecolor{rosso}{cmyk}{0,1,1,0.4}
\definecolor{rossos}{cmyk}{0,1,1,0.55}
\definecolor{rossoc}{cmyk}{0,1,1,0.2}
\definecolor{blu}{cmyk}{1,1,0,0.3}
\definecolor{blus}{cmyk}{1,1,0,0.6}
\definecolor{bluc}{cmyk}{1,1,0,0.1}
\definecolor{verde}{cmyk}{0.92,0,0.59,0.25}
\definecolor{verdec}{cmyk}{0.92,0,0.59,0.15}
\definecolor{verdes}{cmyk}{0.92,0,0.59,0.4}

\renewcommand\&{&}

\def\circa#1{\,\raise.3ex\hbox{$#1$\kern-.75em\lower1ex\hbox{$\sim$}}\,}

\newcommand{\beq}{\begin{equation}}
	\newcommand{\eeq}{\end{equation}}

\newcommand{\bea}{\begin{eqnarray}}
	\newcommand{\eea}{\end{eqnarray}}
\newcommand{\be}{\begin{equation}}
	\newcommand{\ee}{\end{equation}}
\font\tenrsfs=rsfs10 at 12pt
\font\sevenrsfs=rsfs7 at 10 pt
\font\fiversfs=rsfs5
\newfam\rsfsfam
\textfont\rsfsfam=\tenrsfs
\scriptfont\rsfsfam=\sevenrsfs
\scriptscriptfont\rsfsfam=\fiversfs
\def\mathscr#1{{\fam\rsfsfam\relax#1}}

\makeatletter

\def\hhref#1{\href{http://arxiv.org/abs/#1}{arXiv:#1}} 

\newcommand{\doi}[1]{\href{http://dx.doi.org/#1}{[doi]}}

\def\hhref#1{\href{http://arxiv.org/abs/#1}{arXiv:#1}}

\def\art{\@ifnextchar[{\eart}{\oart}}
\def\eart[#1]#2#3#4#5#6{{\rm #2}, {\em #3 \bf #4} {\rm (#6) #5} ({\em #1})}

\def\article{\@ifnextchar[{\earticle}{\oarticle}}
\def\oarticle#1#2#3#4#5#6{{\rm #1}, {\em ``#6''}, {\rm #2 #3 (#5) #4}}
\def\earticle[#1]#2#3#4#5#6#7{{\rm #2}, {\em ``#7''}, {\rm #3 #4 (#6) #5}  [\hhref{#1}]}
\def\hepart[#1]#2{{\rm #2, \em#1}}
\def\heparticle[#1]#2#3{#2, {\em ``#3''} [\hhref{#1}]}

\newcounter{alphaequation}[equation]
\def\thealphaequation{\theequation\hbox to
	0.6em{\hfil\alph{alphaequation}\hfil}}
\def\eqnsystem#1{
	\def\@eqnnum{{\rm (\thealphaequation)}}
	\def\@@eqncr{\let\@tempa\relax \ifcase\@eqcnt \def\@tempa{& & &} \or
		\def\@tempa{& &}\or \def\@tempa{&}\fi\@tempa
		\if@eqnsw\@eqnnum\refstepcounter{alphaequation}\fi
		\global\@eqnswtrue\global\@eqcnt=0\cr}
	\refstepcounter{equation} \let\@currentlabel\theequation \def\@tempb{#1}
	\ifx\@tempb\empty\else\label{#1}\fi
	\refstepcounter{alphaequation}
	\let\@currentlabel\thealphaequation
	\global\@eqnswtrue\global\@eqcnt=0 \tabskip\@centering\let\\=\@eqncr
	$$\halign to \displaywidth\bgroup \@eqnsel\hskip\@centering
	$\displaystyle\tabskip\z@{##}$&\global\@eqcnt\@ne
	\hskip2\arraycolsep\hfil${##}$\hfil& \global\@eqcnt\tw@\hskip2\arraycolsep
	$\displaystyle\tabskip\z@{##}$\hfil
	\tabskip\@centering&\llap{##}\tabskip\z@\cr}
\def\endeqnsystem{\@@eqncr\egroup$$\global\@ignoretrue} \makeatother

\definecolor{fiorentina}{rgb}{.5,0,.5}

\usepackage{tikz-feynman}

\let\oldheadrule\headrule
\renewcommand{\headrule}{\color{white}\oldheadrule}
\usepackage{graphicx} 

\usepackage{orcidlink}

\begin{document}
 \begin{center}
		\boldmath

		{\textbf{\LARGE\color{black!35!blue}
\bf {\LARGE  Defect Induced Heavy Meson   Dynamics   \\  [.4cm] in \\ [.6cm]
          The QCD Conformal Window  
            }}}
            
		\unboldmath
		
		\bigskip\bigskip

	 	{Vigilante~Di~ Risi$^{1,3\,\star}$, Davide~Iacobacci$^{1,3\,\dagger}$,   Francesco~Sannino$^{1,2,3\,^\diamondsuit}$\\[5mm]}

\small{$^1$ Dipartimento di Fisica ``E. Pancini", Università di Napoli Federico II - INFN sezione di Napoli, Complesso Universitario di Monte S. Angelo Edificio 6, via Cintia, 80126 Napoli, Italy}\\
\small{$^2$ Scuola Superiore Meridionale, Largo S. Marcellino, 10, 80138 Napoli NA, Italy}\\
$^3$  Quantum  Theory Center ($\hbar$QTC), Danish-IAS, IMADA, Southern Denmark Univ., Campusvej 55, 5230 Odense M, Denmark\\

		\bigskip\bigskip
		
		\thispagestyle{empty}\large
		{\bf\color{blus} Abstract}
		\begin{quote} 
		\normalsize

 Upon introducing an heavy quark in the perturbative regime of the QCD conformal window we precisely determine the associated heavy meson spectrum and wave functions in terms of the number of light flavours and mass.  We then compute the conformal Isgur-Wise function which is a central quantity in heavy quark physics. We further determine the impact of the residual low energy confining dynamics on the heavy meson spectrum. As a working framework, we adapt the heavy quark effective theory to the perturbative conformal window dynamics. Our work lays the foundations  to  systematically go beyond the infinite mass defect approximation in conformal field theories.

\noindent

\end{quote}
\end{center}
\newpage 
\tableofcontents

\newpage
\section{Introduction} 
\label{intro}

The \emph{conformal window} is the region in the number of flavours versus colours of asymptotically free gauge theories where the theory achieves conformality in the infrared. The phase diagram for QCD and QCD-like theories appeared in \cite{Sannino:2004qp,Dietrich:2006cm}.    At the lower end of the conformal window the theory undergoes a
quantum phase transition from an infrared conformal field theory to a phase characterized by both conformal and chiral symmetry breaking \cite{Miransky:1984ef,Miransky:1996pd}. An exciting possibility is that the loss of conformality could occurr à la Berezinskii-Kosterlitz-Thouless (BKT) \cite{Berezinsky:1970fr,Kosterlitz:1973xp,Kosterlitz:1974nba}.  This happens in two dimensions and was imagined to occur in four dimensions in \cite{Miransky:1996pd, Miransky:1984ef,Appelquist:1996dq}. An alternative and still valid scenario is that the quantum transition is a jumping one \cite{Sannino:2012wy}. A smooth quantum phase transition could lead to  non-conformal physics displaying premonitory signs of quasi conformality observable as  power-law scaling of certain operators \cite{Holdom:1988gs, Holdom:1988gr,Cohen:1988sq}.  The associated near-conformal phase is characterized by the existence of a region of the renormalization group (RG) flow in which the coupling remains nearly constant, signaling the occurrence of \emph{walking dynamics} \cite{Holdom:1988gs,Holdom:1988gr,Cohen:1988sq}. Walking behavior lies at the core of several phenomenological models of dynamical electroweak symmetry breaking e.g. within the technicolor \cite{Weinberg:1975gm, Cohen:1988sq, Appelquist:1999dq, Duan:2000dy, Sannino:2004qp,Dietrich:2005jn,Dietrich:2006cm, Cacciapaglia:2020kgq} and (fundamental partial) composite Goldstone Higgs scenarios \cite{Kaplan:1983fs, Kaplan:1983sm, Cacciapaglia:2014uja, Gripaios:2009pe, Galloway:2010bp, Barnard:2013zea, Sannino:2016sfx, Orlando:2020yii, Bersini:2024twu} as well as epidemiology and population dynamics \cite{10.3389/fams.2021.659580}.

Here, we investigate the perturbative regime
of the quantum chromodynamics  conformal window in presence of a Wilson line defect represented by an heavy quark.  The role of the traditional non-perturbative QCD  \textit{brown muck} (see reference \cite{Isgur:1989vq,Isgur:1990yhj,Grinstein:1991ap} for the origin of the term) is  replaced by its weakly coupled \textit{conformal brown muck} counterpart. We then adapt the Heavy Quark Effective Theory \cite{Neubert:1993mb} framework to provide a counting scheme for the corrections stemming from the finite heavy quark mass. This allows to take into account the {\it wiggle} of the  Wilson line and, de facto, renders the defect dynamical. The impact of defects in conformal field theories is currently an active field of research for both abelian  and non-abelian theories \cite{Cuomo:2021rkm,Cuomo:2022xgw,Aharony:2022ntz,Aharony:2023amq} where the non-interacting fixed points constitute the bulk conformal theory. These studies concentrate on the modification of the bulk dynamics when changing the value of the charge of the Wilson line. 

In our work, the bulk dynamics is constituted by the interacting conformal brown muck. Additionally, we concentrate on its perturbative region  although the formalism can be readily extended to the non-perturbative regime as well. The setup allows us to explore different properties and access different quantum states, such as heavy mesons made by an heavy quark and a light antiquark. Specifically, we precisely determine the  heavy meson spectrum and its wave functions in terms of the number of light flavours and mass.  We consider different light quark mass scenarios and furthermore provide the conformal Isgur-Wise function \cite{Isgur:1989vq,Isgur:1990yhj} which encodes the information of the way the light conformal cloud adjusts in the heavy to heavy quark transitions featuring two distinct super-selected velocities. Last but not least, when providing masses to all the light quarks, the theory develops an exponentially small confining scale due to the residual Yang-Mills dynamics which we take into account for the heavy meson spectrum. 

The paper is structured as follows. In Section~\ref{theoretical} we delineate the theoretical framework for the conformal window in presence of an heavy quark and consider different scenarios for the light quark masses. We further introduce the framework for the heavy quark dynamics. The heavy quark potential and associate  heavy meson bound states and wave functions are computed in Section~\ref{PotentialHeavylight} together with the conformal Isgur-Wise function. The impact of the exponentially suppressed confinement potential is determined in Section~\ref{confpotential}. We offer our outlook and conclusions in Section~\ref{conclusions}. 


\section{Theoretical Framework}
\label{theoretical}

We consider  an asymptotically free $SU(N)$ gauge theory with $N_F$ massless fermions featuring a  Banks and Zaks (BZ) \cite{Banks:1981nn} infrared perturbative fixed point (IRFP) obtained by  adjusting the ratio $N/N_F$. The occurrence of the IRFP appears first at the two-loop level. The
analysis at higher loops has been performed in \cite{Mojaza:2012zd,Litim:2015iea,Cacciapaglia:2020kgq}. The generalization of the BZ work, beyond the fundamental representation, started first for the two-index fermion theories in \cite{Sannino:2004qp} and extended to all relevant representations in \cite{Dietrich:2006cm} with orthogonal and simplectic gauge-fermion theories investigated in \cite{Sannino:2009aw} and exceptional groups in \cite{Mojaza:2012zd}. These studies were instrumental to the discovery of four-dimensional asymptotically safe gauge theories in \cite{Litim:2015iea}.  Although we focus our attention on asymptotically free $SU(N)$ gauge-fermion theories the methodology is adaptable to other gauge groups and matter representations featuring perturbative fixed points.

\subsection{Perturbative infrared conformal dynamics setup}
\label{subsec:conformal_window}

At the two loop level the beta function $\beta(\alpha_s)$ for the coupling $\alpha_s = g_s^2/{4\pi}$ reads: 
\begin{equation}
    \label{beta_func}
    \mu^2 \frac{d}{d \mu^2} \Big(\frac{\alpha_s}{2\pi}\Big) = - \frac{1}{2}\beta_0 \Big(\frac{\alpha_s}{2\pi}\Big)^2- \frac{1}{4} \beta_1 \Big(\frac{\alpha_s}{2\pi}\Big)^3  \ ,
\end{equation}
where 
\begin{equation}
    \begin{aligned}
       & \beta_0= \frac{11}{3}C_A -\frac{4}{3} T(r) N_F\ , \\
       & \beta_1= \frac{34}{3} C_A^2-4 \Big( \frac{5}{3}C_A+C_r\Big)T(r) N_F \ .
    \end{aligned}
\end{equation}
Here, $T(r)$ is the trace normalization factor of the generators of the $SU(N)$ gauge group in the representation $r$: $T^{\,r}_a T^{\,r}_b=T(r) \delta_{ab}$; $C_A=N$ is the quadratic Casimir in the adjoint representation, while $C_r$ denotes the quadratic Casimir in the representation $r$. \\
The trace normalization and the quadratic Casimir are related by $C_r d(r) = T(r) d(A)$, where $d(r)$ is the dimension of the representation $r$. For example, for QCD the number of colors is $N=3$, and the $N_F$ massless Dirac fermions  transform according to the fundamental representation of the gauge group.   The loss of asymptotic freedom occurs for
\begin{equation}
\label{eq:loss_of_asymptotic_freedom}
   N_F^{\rm AF}= \frac{11}{4} \frac{C_A}{T(r)}= \frac{11}{4}\frac{d(A) C_A}{d(r) C_r}  \ .
\end{equation}
For $N_F$ close to  $N_F^{\rm AF}$ from below, the theory develops a BZ-like IRFP where the second coefficient of the beta function $\beta_1$ is negative. The fixed point is obtained for: 
\begin{equation}
\label{alphastar}
    \alpha_s^*=-4\pi \beta_0/\beta_1 = -  \frac{44}{3} \pi \, C_A  \frac{\epsilon}{\beta_1(N_F^{AF})}=  \frac{44 \pi }{21 C_A +33 C_r}  \,\epsilon +\mathcal{O}(\epsilon^2) \ ,
\end{equation}
where $N_F =N_F^{\rm AF} (1 - \epsilon)$  and we assumed the number of flavors to be real numbers so that one can formally take $\epsilon \ll 1$.  
Increasing $\epsilon$, the fixed point value increases and eventually the interactions are sufficiently strong to trigger dynamical chiral symmetry breaking and consequently loose conformality.  Uncovering the non-perturbative physics occurring at the lower end of the conformal window is a long-standing problem  \cite{Sannino:2009za,Cacciapaglia:2020kgq, Sannino:2004qp, Dietrich:2006cm, Kim:2020yvr, Ryttov:2010iz, Pagels:1974se, Ryttov:2007cx, Ryttov:2016hdp, Hasenfratz:2019dpr, Appelquist:2009ty, Appelquist:2007hu, Appelquist:2011dp, Chiu:2018edw, Pica:2010xq, Hasenfratz:2023wbr, Chiu:2018edw, Appelquist:2011dp}.  We shall not be concerned here with these issues but rather focus on the exciting possibility to study novel (near) conformal properties of the BZ theory in the presence of an heavy quark. 

\begin{figure}
    \centering
    \begin{tikzpicture}
[x=0.6pt,y=0.6pt,yscale=-0.9,xscale=0.9]

\draw [color={rgb, 255:red, 0; green, 0; blue, 0 }  ,draw opacity=1 ][line width=0.75]  (46,232.2) -- (573.33,232.2)(98.73,18) -- (98.73,256) (566.33,227.2) -- (573.33,232.2) -- (566.33,237.2) (93.73,25) -- (98.73,18) -- (103.73,25)  ;
\draw [color={rgb, 255:red, 208; green, 2; blue, 27 }  ,draw opacity=1 ][line width=1.5]    (238.33,130) .. controls (256.45,130.46) and (268.7,132.97) .. (277.67,136.89) .. controls (295.76,144.8) and (300.57,158.49) .. (313.44,172.9) .. controls (326.51,187.52) and (347.87,202.9) .. (399.8,213.75) .. controls (435.31,221.17) and (485.12,226.48) .. (556.33,228) ;
\draw [color={rgb, 255:red, 208; green, 2; blue, 27 }  ,draw opacity=1 ][line width=1.5]    (98.33,129) -- (168.32,129.5) -- (179.34,129.58) -- (184.35,129.61) -- (208.33,129.79) -- (238.33,130) ;
\draw [color={rgb, 255:red, 0; green, 0; blue, 0 }  ,draw opacity=1 ][line width=0.75]  [dash pattern={on 4.5pt off 4.5pt}]  (298.33,20.67) -- (299.33,231.67) ;

\draw (559,244.4) node [anchor=north west][inner sep=0.75pt]    {$\mu $};
\draw (68,14.4) node [anchor=north west][inner sep=0.75pt]    {$\alpha _{s}$};
\draw (70,117.4) node [anchor=north west][inner sep=0.75pt]    {$\alpha _{s}^{*}$};
\draw (292,237) node [anchor=north west][inner sep=0.75pt]    {$\Lambda _{RGI}$};

\end{tikzpicture}
 \caption{The  running of the coupling constant when a BZ infrared fixed point arises. At the scale $\Lambda_{RGI}$ the RG-flow enters in the domain of the IRFP where the theory displays a Coulomb-like behavior.}
    \label{fig:BZ-running}
\end{figure}
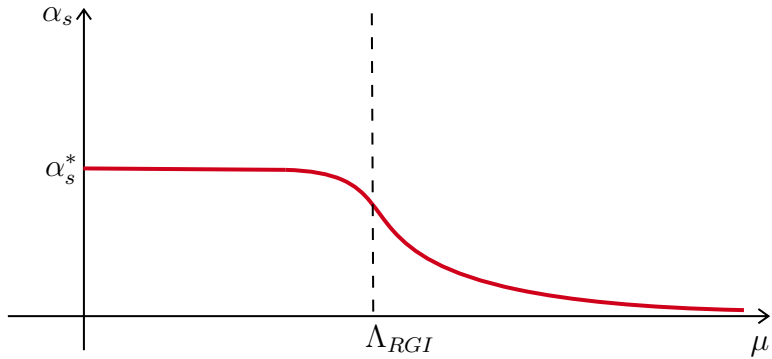

Solving for the two-loop  running of the coupling  $\alpha_s$ between the asymptotically free limit and the interacting IRFP one obtains \cite{Litim:2015iea,DiGiustino:2021nep}: 
\begin{equation}
\label{eq:solution-lambert-equation}
    \alpha_s= \frac{\alpha_s^*}{1+W(z)} \ ,
\end{equation}
where $W(z)$ is the Lambert function
\begin{equation}
\label{eq:lambert-equation}
    We^W=z \ ,
\end{equation} 
with 
\begin{equation}
\label{eq:z-lambert-equation}
    z= e^{\frac{\alpha_s^*}{\alpha_{s_{0}}}-1}\Bigg(\frac{\alpha_s^*}{\alpha_{s_0}}-1\Bigg) \Bigg(\frac{\mu_0}{\mu}\Bigg)^{ -B \, \alpha_s^* } \ .  
\end{equation}
We have defined $B = \beta_0/2\pi$. The initial condition for the coupling $\alpha_{s_0}$ must be smaller than $\alpha_s^\ast$ to ensure that we pick the branch of the beta function connecting the asymptotically free fixed point with the IRFP. For a more general discussion see \cite{DiGiustino:2021nep}. 

The theory yields a new dynamically-generated scale  $\Lambda_{RGI}$ which is also renormalization group invariant. It marks the transition from the asymptotically free dominated UV physics to the interacting IRFP regime. Following \cite{Mojaza:2012zd,Litim:2015iea}, to determine  $\Lambda_{RGI}$ we first introduce  
\begin{equation}
    \theta(\alpha_s)=\partial_{\alpha_s}\beta=-2\beta_0\frac{\alpha_s}{(2\pi)^2}\bigg(1-\frac{3}{2}\frac{\alpha_s}{\alpha_s^*}\bigg)  \ , 
    \end{equation}
which at both fixed points reduces to the scaling exponents
\begin{align}
    &\theta(\alpha_s\rightarrow 0^+)=-2\beta_0\frac{\alpha_s}{(2\pi)^2}\rightarrow 0^- \ ,\\
    &\theta(\alpha_s\rightarrow\alpha_s^*)=\beta_0 \frac{\alpha_s^*}{(2\pi)^2}>0 . 
\end{align}
We therefore have that, as the theory departs from the gaussian UV fixed point,  it approaches the interacting IRFP at low energies.  The  $\Lambda_{RGI}$ scale is mathematically defined as the one for which $\theta$ changes sign, \textit{i.e.} $\alpha_s=\frac{2}{3}\alpha^*_s$.  Its expression is therefore  derived from Eq.~\eqref{eq:solution-lambert-equation} and Eq.~\eqref{eq:z-lambert-equation}: 

\begin{equation}
    \Lambda_{RGI}=\mu \exp[\bigg(\ln 2 -1/2+\frac{\alpha_s^*-\alpha}{\alpha}\bigg)\frac{1}{\theta(\alpha_s^*)}]\abs{\frac{\alpha^*_s-\alpha_s}{\alpha_s}}^{\frac{1}{\theta(\alpha_s^*)}}.
\end{equation}

In terms of small deviations $\delta \alpha=\alpha_s-\alpha_s^*$ from the IRFP it reduces to:

\begin{equation}
    \Lambda_{RGI}=\mu \exp[(\ln 2 -1/2)\frac{1}{\theta(\alpha_s^*)}]\abs{1-\frac{\alpha_s}{\alpha^*_s}}^{\frac{1}{\theta(\alpha_s^*)}} \ .
\end{equation}
That $\Lambda_{RGI}$ is actually a RG-invariant scale \cite{Litim:2015iea} can be checked explicitly by showing that its variation with respect to $\mu$ vanishes upon using Eq.~\eqref{eq:solution-lambert-equation} and Eq.~\eqref{eq:z-lambert-equation}. 

The previous discussion can be formally extended to integer values of $N_F$ fundamental fermions provided we consider the limit of both an  infinite number of flavours and colors such that the ratio $N_F/N$ is fixed.  Although this is the limit we have in mind, to make immediate contact with physical and lattice applications we consider, in the following, three colors with a number of fundamental quarks near the loss of asymptotic freedom.  

\subsection{Heavy Quark Dynamics} 

\begin{figure}[h]
    \centering
 \tikzset{every picture/.style={line width=0.75pt}} 

\begin{tikzpicture}[x=0.40pt,y=0.40pt,yscale=-1,xscale=1]

\draw  (51.55,211.27) -- (584.89,213.04)(105.52,20.94) -- (104.82,232.61) (577.91,208.01) -- (584.89,213.04) -- (577.88,218.01) (100.5,27.92) -- (105.52,20.94) -- (110.5,27.95)  ;
\draw [color={rgb, 255:red, 208; green, 2; blue, 27 }  ,draw opacity=1 ][line width=1.5]    (243.33,81) .. controls (288.15,81.08) and (318.89,88.21) .. (342.54,99.05) .. controls (350.19,102.56) and (357.09,106.46) .. (363.49,110.63) .. controls (419.9,147.41) and (437.02,205.71) .. (576.33,208.67) ;
\draw [color={rgb, 255:red, 208; green, 2; blue, 27 }  ,draw opacity=1 ][line width=1.5]    (105.33,79.33) -- (200.33,80.48) -- (210.34,80.6) -- (243.33,81) ;
\draw  [dash pattern={on 0.84pt off 2.51pt}]  (276.33,150.67) .. controls (359.77,151.46) and (399.29,164.98) .. (427.33,175.33) .. controls (455.38,185.68) and (484.73,205.85) .. (576.33,208.67) ;
\draw  [dash pattern={on 0.84pt off 2.51pt}]  (105.33,148.67) -- (276.33,150.67) ;
\draw  [dash pattern={on 4.5pt off 4.5pt}]  (502.33,19) -- (502.33,212) ;
\draw  [dash pattern={on 4.5pt off 4.5pt}]  (457.33,19.67) -- (458,213) ;

\draw (60,25.4) node [anchor=north west][inner sep=0.35pt]    {$\alpha _{s}$};
\draw (558,222.4) node [anchor=north west][inner sep=0.35pt]    {$\mu $};
\draw (486,222.4) node [anchor=north west][inner sep=0.35pt]    {$M$};
\draw (30,136.4) node [anchor=north west][inner sep=0.35pt]  [font=\tiny]  {${\textstyle \alpha _{s}^{*}( N_{F})}$};
\draw (2,70.4) node [anchor=north west][inner sep=0.35pt]  [font=\tiny]  {$\alpha _{s}^{*}( N_{F} -1)$};
\draw (417,220.4) node [anchor=north west][inner sep=0.75pt]  [font=\normalsize]  {$\Lambda _{RGI}$};

\end{tikzpicture}
\begin{tikzpicture}[x=0.38pt,y=0.38pt,yscale=-1,xscale=1]

\draw  (50,229.3) -- (546.33,229.3)(99.63,42) -- (99.63,249) (539.33,224.3) -- (546.33,229.3) -- (539.33,234.3) (94.63,49) -- (99.63,42) -- (104.63,49)  ;
\draw [color={rgb, 255:red, 208; green, 2; blue, 27 }  ,draw opacity=1 ][line width=1.5]    (355.33,154.33) .. controls (466.33,156.33) and (370.33,221.33) .. (533.33,224) ;
\draw [color={rgb, 255:red, 208; green, 2; blue, 27 }  ,draw opacity=1 ][line width=1.5]    (99.33,106.33) -- (138.34,106.56) -- (156.33,106.67) ;
\draw [color={rgb, 255:red, 208; green, 2; blue, 27 }  ,draw opacity=1 ][line width=1.5]    (156.33,106.67) .. controls (192.33,111.33) and (145.33,151.33) .. (268.33,154.33) ;
\draw [color={rgb, 255:red, 208; green, 2; blue, 27 }  ,draw opacity=1 ][line width=1.5]    (268.33,154.33) -- (355.33,154.33) ;
\draw  [dash pattern={on 4.5pt off 4.5pt}]  (270.33,53.33) -- (271,230) ;
\draw  [dash pattern={on 0.84pt off 2.51pt}]  (98.33,154.33) -- (268.33,154.33) ;
\draw  [dash pattern={on 4.5pt off 4.5pt}]  (418.33,54.33) -- (421,231) ;

\draw (520,237.4) node [anchor=north west][inner sep=0.35pt]    {$\mu $};
\draw (55,51.4) node [anchor=north west][inner sep=0.35pt]    {$\alpha _{s}$};
\draw (253,237.4) node [anchor=north west][inner sep=0.35pt]    {$M$};
\draw (-9,96.4) node [anchor=north west][inner sep=0.35pt]  [font=\tiny]  {$\alpha _{s}^{*}( N_{F} -1)$};
\draw (25,145.4) node [anchor=north west][inner sep=0.35pt]  [font=\tiny]  {$\alpha _{s}^{*}( N_{F})$};
\draw (390,235.4) node [anchor=north west][inner sep=0.35pt]    {$\Lambda _{RGI}$};

\end{tikzpicture}
\caption{The strong coupling running (solid red curve) with a heavy quark above (left) or below (right) the $\Lambda_{RGI}$ scale of the theory with $N_F$ massless quarks (dashed lines). }
\label{fig:BZ-heavymass-running}
\end{figure}
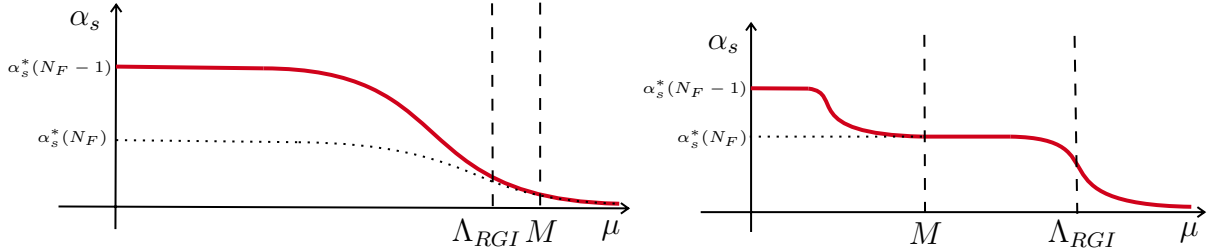

Let the $N_F$-th quark $Q$ acquire mass $M >  \Lambda_{RGI}$ with $\Lambda_{RGI}$ the RGI invariant scale for $N_F$ massless quarks (i.e. the would be RGI  scale before introducing $M$). The cartoon for the associated  running of the coupling (solid red curve) is given in Fig.~\ref{fig:BZ-heavymass-running} left panel. The dashed black curve represents the running for $N_F$ massless quarks for which $\Lambda_{RGI}$ is defined. Since we employ a mass-independent renormalization scheme, the running above   $M$ is given by the theory with $N_F$ massless degrees of freedom and below it by $N_F-1$ massless quarks with a matching of the couplings at the $M$ energy scale. In the right panel of Fig.~\ref{fig:BZ-heavymass-running} we present the case in which $M < \Lambda_{RGI}$. Clearly in the infrared we always arrive at the same IRFP determined by the theory with $N_F-1$ massless quarks indicated by $\alpha_s^\ast (N_F-1)$. Inevitably $\alpha_s^\ast (N_F-1) > \alpha_s^\ast (N_F)$. 

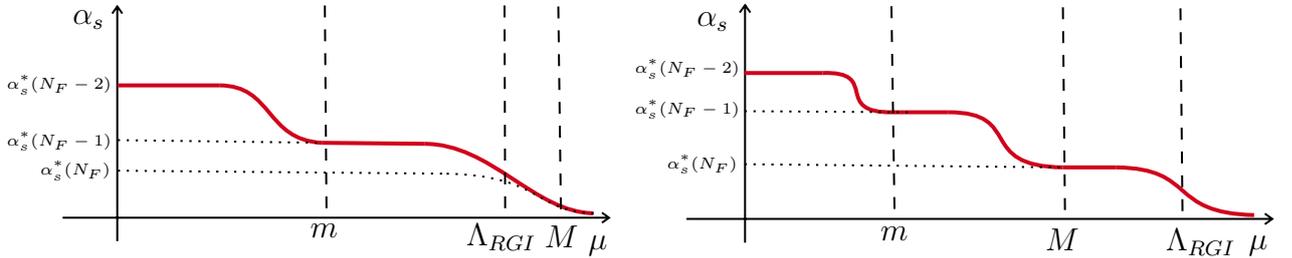
\begin{figure}[h]
    \centering
    \tikzset{every picture/.style={line width=0.75pt}} 
\begin{tikzpicture}[x=0.38pt,y=0.38pt,yscale=-1,xscale=1]

\draw  (50,272.7) -- (588.33,272.7)(103.83,63) -- (103.83,296) (581.33,267.7) -- (588.33,272.7) -- (581.33,277.7) (98.83,70) -- (103.83,63) -- (108.83,70)  ;
\draw [color={rgb, 255:red, 208; green, 2; blue, 27 }  ,draw opacity=1 ][line width=1.5]    (305.33,198.33) -- (406.33,199.33) ;
\draw [color={rgb, 255:red, 208; green, 2; blue, 27 }  ,draw opacity=1 ][line width=1.5]    (406.33,199.33) .. controls (472.33,199.33) and (514.33,267.33) .. (573,268.33) ;
\draw [color={rgb, 255:red, 208; green, 2; blue, 27 }  ,draw opacity=1 ][line width=1.5]    (204.33,141.33) .. controls (257.67,141.33) and (249.33,195.33) .. (305.33,198.33) ;
\draw [color={rgb, 255:red, 208; green, 2; blue, 27 }  ,draw opacity=1 ][line width=1.5]    (104.33,141.33) -- (204.33,141.33) ;
\draw  [dash pattern={on 0.84pt off 2.51pt}]  (428.33,229) .. controls (522.33,231) and (514.33,266.33) .. (573,268.33) ;
\draw  [dash pattern={on 0.84pt off 2.51pt}]  (104.33,226) -- (428.33,229) ;
\draw  [dash pattern={on 4.5pt off 4.5pt}]  (538.33,63) -- (541,274) ;
\draw  [dash pattern={on 4.5pt off 4.5pt}]  (485.33,62) -- (486,273) ;
\draw  [dash pattern={on 4.5pt off 4.5pt}]  (308.33,62) -- (310,272) ;
\draw  [dash pattern={on 0.84pt off 2.51pt}]  (104.33,195.67) -- (305.33,198.33) ;

\draw (522,278) node [anchor=north west][inner sep=0.75pt]   [align=left] {$\displaystyle M$};
\draw (445,276) node [anchor=north west][inner sep=0.75pt]   [align=left] {$\displaystyle \Lambda _{RGI}$};
\draw (291,276) node [anchor=north west][inner sep=0.75pt]   [align=left] {$\displaystyle m$};
\draw (565,288) node [anchor=north west][inner sep=0.75pt]   [align=left] {$\displaystyle \mu $};
\draw (57,64) node [anchor=north west][inner sep=0.75pt]   [align=left] {$\displaystyle \alpha _{s}$};
\draw (-8,128) node [anchor=north west][inner sep=0.75pt]  [font=\tiny] [align=left] {$\displaystyle \alpha _{s}^{*}( N_{F} -2)$};
\draw (25,213) node [anchor=north west][inner sep=0.75pt]  [font=\tiny] [align=left] {$\displaystyle \alpha _{s}^{*}( N_{F})$};
\draw (-8,184) node [anchor=north west][inner sep=0.75pt]  [font=\tiny] [align=left] {$\displaystyle \alpha _{s}^{*}( N_{F} -1)$};
\end{tikzpicture}
\tikzset{every picture/.style={line width=0.75pt}} 
\begin{tikzpicture}[x=0.40pt,y=0.40pt,yscale=-1,xscale=1]

\draw  (46,271.6) -- (594.33,271.6)(100.83,70) -- (100.83,294) (587.33,266.6) -- (594.33,271.6) -- (587.33,276.6) (95.83,77) -- (100.83,70) -- (105.83,77)  ;
\draw [color={rgb, 255:red, 208; green, 2; blue, 27 }  ,draw opacity=1 ][line width=1.5]    (448.33,223) .. controls (524.33,224) and (495.33,266) .. (577.33,268) ;
\draw [color={rgb, 255:red, 208; green, 2; blue, 27 }  ,draw opacity=1 ][line width=1.5]    (398.33,223) -- (448.33,223) ;
\draw [color={rgb, 255:red, 208; green, 2; blue, 27 }  ,draw opacity=1 ][line width=1.5]    (291.33,171) .. controls (362.33,171) and (316.33,222) .. (398.33,223) ;
\draw [color={rgb, 255:red, 208; green, 2; blue, 27 }  ,draw opacity=1 ][line width=1.5]    (239.33,171) -- (291.33,171) ;
\draw [color={rgb, 255:red, 208; green, 2; blue, 27 }  ,draw opacity=1 ][line width=1.5]    (173.33,134) .. controls (228.33,132) and (181.33,172) .. (239.33,171) ;
\draw [color={rgb, 255:red, 208; green, 2; blue, 27 }  ,draw opacity=1 ][line width=1.5]    (100.33,134) -- (173.33,134) ;
\draw  [dash pattern={on 0.84pt off 2.51pt}]  (100.33,170) -- (256.33,171) ;
\draw  [dash pattern={on 4.5pt off 4.5pt}]  (237.83,71) -- (240.83,271) ;
\draw  [dash pattern={on 4.5pt off 4.5pt}]  (398.33,72) -- (400.33,271) ;
\draw  [dash pattern={on 4.5pt off 4.5pt}]  (508.33,73) -- (510.33,273) ;
\draw  [dash pattern={on 0.84pt off 2.51pt}]  (100.33,220) -- (398.33,223) ;

\draw (53,74) node [anchor=north west][inner sep=0.75pt]   [align=left] {
$\alpha_{s}$};
\draw (569.33,285) node [anchor=north west][inner sep=0.75pt]   [align=left] {
$\mu $};
\draw (-5,118) node [anchor=north west][inner sep=0.75pt]  [font=\tiny] [align=left] {$\displaystyle \alpha _{s}^{*}( N_{F} -2)$};
\draw (-5,156) node [anchor=north west][inner sep=0.75pt]  [font=\tiny] [align=left] {$\displaystyle \alpha _{s}^{*}( N_{F} -1)$};
\draw (225,276) node [anchor=north west][inner sep=0.75pt]   [align=left] {$\displaystyle m$};
\draw (379,277.67) node [anchor=north west][inner sep=0.75pt]   [align=left] {$\displaystyle M$};
\draw (492,277.67) node [anchor=north west][inner sep=0.75pt]   [align=left] {$\displaystyle \Lambda _{RGI}$};
\draw (25,208) node [anchor=north west][inner sep=0.75pt]  [font=\tiny] [align=left] {$\displaystyle \alpha _{s}^{*}( N_{F})$};

\end{tikzpicture}

    \caption{Running of $\alpha_s$ (solid red curve) when giving a mass $m \ll M$ to only one quark, with $M>\Lambda_{RGI}$ and $M<\Lambda_{RGI}$ in the left and right panels respectively. The IRFP is $\alpha_s^*(N_F-2)$. The dashed lines indicate the running for $N_F$ and $N_F-1$ active flavors.}
    \label{fig:one-light-quark}
\end{figure}

If we further give a mass $m\ll \Lambda_{RGI}$ (or $m\ll M$ for $M<\Lambda_{RGI}$) to just one of the light flavours  the $N_F-2$ theory can still be conformal and the coupling runs to a new IRFP $\alpha_s^*(N_F-2)$ as shown in  Fig.~\ref{fig:one-light-quark}.  At infinite number of colors and flavours this latter fixed point is still within the perturbative regime. For two massive quarks the conformal window is depicted in Sec. \ref{subsec:conformal_window}. 
 { When giving mass to two of them, the IR theory for three colors features a maximum number of massless flavors $n^{max}_f=N_F^{AF}-2=14.5$, and the theory looses asymptotic freedom for $N_F^{AF}=16.5$ from Eq. \eqref{eq:loss_of_asymptotic_freedom}. In this case, for example, we would have $\epsilon \sim 2/16.5 \approx 0.12$ and $\alpha_s^\ast\approx 0.2$. One can also consider the case in which the heavy quark corresponds to a 17th flavour and in this case above the mass of the heavy quark one looses asymptotic freedom while gaining access to a smaller value of the IR fixed point.

\section{Heavy Meson Spectrum and Dynamics}
\label{PotentialHeavylight}
To determine the spectrum composed by a light antiquark $\Bar{q}$ and an heavy quark $Q$ in the dynamical regimes discussed above we consider solutions of the Dirac equation for the light antiquark in the non-abelian potential generated by the heavy quark. We expect  the bound states to form in the region below the mass $m$ of the light quark, where  $\alpha_s$ is fixed at the value $\alpha_s^*(N_F-2)$ as depicted in \cref{fig:one-light-quark}.\\
In QCD, the typical size of a bound state is of the order of $R_{\text{had}}\sim \Lambda_{\text{QCD}}^{-1}$ \cite{Neubert:1993mb}. This scale corresponds to the regime where the coupling constant becomes strong, resulting in non-perturbative phenomena such as confinement. A quark $Q$ is called heavy if its mass $M \gg \Lambda_{\text{QCD}}$ and one can build an effective theory which allows to facilitate the study of the properties of heavy hadrons, such as their spectroscopy and decay amplitudes.
However, the dynamics of these bound states cannot be fully resolved, as non-perturbative features of QCD persist in the computation of the observables, making it impossible to determine them from first principles.\\
In the (near) conformal field theory the lightest relevant energy scale is given by $m$. Therefore a quark will be considered heavy if its mass $M \gg m$ and the HQET approach can still be employed. Notably, the great advantage of our framework is that it allows the investigation of the properties of bound states using ordinary perturbation theory. By tuning the number of flavors, $N_F$, and the number of colors, $N$, we can ensure that the coupling constant remains sufficiently small.

\subsection{The Heavy Quark Potential}

\label{sec:the-heavy-quark-effective-theory}
In ordinary QCD, the dynamics of an heavy quark bounded to a light antiquark can be described in the framework of the HQET. The key observation is that the heavy quark velocity $v$ is superselected and its field $Q(x)$ can be decomposed into a large and a small component \cite{Neubert:1993mb}
\begin{align}
	\label{eqn:masssub}
	h_v(x)&=e^{iMv\cdot x}P_+Q(x) \ , \notag \\
 H_v(x)&=e^{iMv\cdot x}P_-Q(x)\,,
\end{align}
where $P_{\pm} = (1 \pm \vslash)/2$ are the Dirac projection operators.
In terms of $h_v$ and $H_v$ the QCD Lagrangian takes the following form:
\begin{equation}
	\label{eqn:fullL}
	\mathcal{L}_{\text{QCD}} 
	= \Bar{h}_v i \vcD h_v + \Bar{h}_v i \Dslash_\perp H_v
	+ \Bar{H}_v i \Dslash_\perp h_v - \Bar{H}_v (i \vcD + 2M) H_v\,,
\end{equation}
where $D^\mu=\partial_{\mu}-ig_sT^a$ is the gauge covariant derivative of QCD, 
and $D^\mu_\perp = D^\mu - (\vcD)v^\mu$ denotes the transverse derivative orthogonal to the heavy quark velocity.

The small component $H_v$ describes fluctuations with twice the heavy quark mass $M$ and can be integrated out through the equation of motions and arrive at the effective Lagrangian
%
\begin{equation}
	\label{eqn:eftL}
	\mathcal{L}_{\text{HQET}} = \Bar{h}_v i \vcD h_v +  \Bar{h}_v i\Dslash_\perp \frac{1}{i \vcD + 2M}  i\Dslash_\perp h_v\,,
\end{equation}
which admits a derivative expansion in $i D /2M$. At leading order (LO) in the heavy quark expansion (HQE), Eq. \eqref{eqn:eftL} reduces to 
\begin{equation}
    \mathcal{L}_{\text{HQET}}= \Bar{h}_v i \vcD h_v +\mathcal{O}(1/M),
\end{equation}
describing an on-shell quark moving with velocity $v$.

Allowing for a large but finite mass $M$, the heavy quark starts wiggling along its straight trajectory with a residual momentum $p$ due to the exchange of soft gluons (below the mass $m$ of the light antiquark) with the conformal cloud.  We can take this behaviour into account  by considering the NLO terms of the HQET lagrangian:
\begin{equation}
     \mathcal{L}_{\text{HQET}}= \Bar{h}_v i \vcD h_v +\frac{1}{2M}\Bar{h}_v(iD_{\perp})^2h_v+\frac{g_s}{4M}\Bar{h}_v\sigma_{\alpha\beta}G^{\alpha\beta}h_v+\mathcal{O}(1/4M^2),
\end{equation}
where $\sigma_{\alpha\beta}=\frac{i}{2}[\gamma_{\alpha},\gamma_{\beta}]$ and $ig_sG^{\alpha\beta}=[D^{\alpha},D^{\beta}]$.

The full heavy quark field, in terms of the small and large components, reads 
\begin{equation}
    Q(x)=e^{-iMv\cdot x}(h_v(x)+H_v(x))=e^{-iMv\cdot x}\bigg(1+\frac{1}{i v\cdot D +2M-i\epsilon}i\Dslash_{\perp}\bigg)h_v(x)
\end{equation}
and at NLO in the HQE reduces to
\begin{equation}
\label{eq:heavy-quark-field-expansion}
    Q(x)=e^{-iMv\cdot x}\bigg(1+\frac{i\slashed D_{\perp}}{2M}\bigg)h_v(x)+\mathcal{O}\bigg(\frac{1}{4M^2}\bigg)\ .
\end{equation}
By virtue of Eq. \eqref{eq:heavy-quark-field-expansion} the heavy quark spinor can be substituted by 
\begin{equation}
    u_{Q}(p,s)=\bigg(1+\frac{\slashed{p}-v\cdot p}{2M}\bigg)u_h(v,s)+\mathcal{O}\bigg(\frac{1}{4M^2}\bigg) \ ,
\end{equation}
where
\begin{equation}
\label{HQ_spinor}
    u_h(v,\,s)= \sqrt{\frac{1+v_E}{2 v_E}}\mqty(\chi_s \\ \frac{\boldsymbol{\sigma}\cdot\boldsymbol{v}}{1+v_E}\chi_s) \ , \quad v_E=\sqrt{1+\abs{\boldsymbol{v}}^2} 
\end{equation}
is the heavy-quark spinor in the exact infinite mass limit. The corresponding NLO heavy quark current is

\begin{equation}
\label{eq:heavy_quark_current}
    \Bar{u}_h(v)\bigg(v^\mu+\frac{(2p-k)^\mu-v\cdot(2p-k)v^{\mu}}{2 M}+\frac{[\gamma^\mu, \slashed{k}-v\cdot k \slashed{v}]}{4 M}\bigg)u_h(v) \ ,
\end{equation}
where $p$ and $p-k$ are the incoming and outcoming residual momenta of the heavy quark when a gluon of momentum $k$ is exchanged.\\
Following the procedure outlined in \cite{Lucha:1991jy,Lucha:1995zv}, we compute the potential generated by the heavy quark from the S-matrix of the quark-antiquark scattering. Using Eq.~\eqref{eq:heavy_quark_current} first-order mass corrections are automatically taken into account.  

In the heavy quark rest frame $v^{\mu}=(1,0,0,0)$ and in the large $M$ limit we can take $k^2=-\boldsymbol{k}^2$. The $T$-matrix element is given by
\begin{equation}
    \bra{p-k,q_f}T\ket{p,q_i}=\frac{1}{(2\pi)^3}\frac{m}{\sqrt{E_{q_i}E_{q_f}}}\frac{4}{3}\frac{g_s^{*2}}{\boldsymbol{k}^2}\bigg[\Bar{v} \gamma^0 v \Bar{u}_h u_h+\Bar{v} \gamma^i v \Bar{u}_h\bigg(\frac{P_i}{M}-\frac{i \epsilon_{ijk}k^j \sigma^k}{2M}\bigg)u_h\bigg] \ ,
\end{equation}
where $q_i$ and $q_f$ are the initial and final light antiquark momenta, and $\boldsymbol{P}=\boldsymbol{p}-\boldsymbol{k}/2$.
Following \cite{Morishita:1987rs}, we make the substitutions 
\begin{align}
   &  \langle \boldsymbol{P}\rangle =\Bar{u}_{h}\boldsymbol{P} u_h \rightarrow \boldsymbol{P}\ , \qquad \langle \boldsymbol{\sigma}\rangle =\Bar{u}_h \boldsymbol{\sigma}u_h \rightarrow \boldsymbol{\sigma}_Q \ .
\end{align}
Then, the $T$-matrix element becomes
\begin{equation}
\label{Tmatrix}
     \bra{p-k,q_f}T\ket{p,q_i}=\frac{1}{(2\pi)^3}\frac{m}{\sqrt{E_{q_i}E_{q_f}}}\frac{4}{3}\frac{g_s^{*2}}{\boldsymbol{k}^2} v^{\dagger}\bigg[1+\boldsymbol{\alpha}_D\cdot\frac{\boldsymbol{P}}{M}+\boldsymbol{\alpha}_D \cdot \frac{\boldsymbol{\sigma}_Q\times \boldsymbol{k}}{2 M}\bigg]v \ ,
\end{equation}
where $\boldsymbol{\alpha}_D$ and $\beta_D$ are the Dirac matrices. From Eq. \eqref{Tmatrix}, we recognise that the potential in momentum-space is given by
\begin{equation}
    \Tilde{V}(\boldsymbol{k})=\frac{4}{3}\frac{g_s^{*2}}{\boldsymbol{k}^2}\bigg(1+\boldsymbol{\alpha}_D\cdot\frac{\boldsymbol{P}}{M}+\boldsymbol{\alpha}_D \cdot \frac{\boldsymbol{\sigma}_Q\times \boldsymbol{k}}{2 M}\bigg) \ .
\end{equation}
By performing the Fourier transformation, we find the potential in the configuration space
\begin{align}
    V(\boldsymbol{r})=&-\frac{1}{(2\pi)^3}\int d\boldsymbol{k}e^{-i\boldsymbol{k}\cdot\boldsymbol{r}}\frac{4}{3}\frac{g_s^{*2}}{\boldsymbol{k}^2}\bigg(1+\boldsymbol{\alpha}_D\cdot\frac{\boldsymbol{P}}{M}+\boldsymbol{\alpha}_D \cdot \frac{\boldsymbol{\sigma}_Q\times \boldsymbol{k}}{2 M}\bigg) \notag \\
    =&-\frac{4}{3}\frac{\alpha_s^{*}}{r}\bigg(1+\boldsymbol{\alpha}_D\cdot\frac{\boldsymbol{P}}{M}\bigg)+\frac{2}{3M}\frac{\alpha_s^{*}}{r^2}\boldsymbol{\alpha}_D \cdot \boldsymbol{\sigma}_{Q}\times \boldsymbol{n} \ ,
\end{align}
where $\boldsymbol{n}\equiv \boldsymbol{r}/r$. 
The leading term in the inverse of the quark mass is heavy quark mass independent and corresponds to the standard potential for an infinite heavy source corresponding also to the defect term introduced in conformal field theories \cite{Aharony:2022ntz,Aharony:2023amq}. Although we shall stop to the leading order in the inverse of the heavy quark mass it is clear that we one can systematically investigate the dynamics of the defect that, at subleading order in the inverse mass expansion, becomes dependent on its quantum numbers like spin and details of its gauge properties.  

 Expanding upon our framework, one could be interested in exploring QCD radiative corrections associated with the interaction between a light antiquark and a heavy quark potential. These corrections stem from the quark self-energy, vertex correction, and vacuum polarization with the well-known Feynman diagrams given in \cref{fig:feynmanloop} as described in \cite{Ryder:1985wq,Muta:1998vi}.
The impact of these corrections, once taken into account for the potential, affect the bound state energies at the $\mathcal{O}(\alpha_s^5)$ level as it is well known already for QED \cite{Berestetskii:1982qgu,greiner2003quantum}.  Consequently, they are beyond the scope of this preliminary investigation.
We are now ready to investigate the bound states of the theory. 
\begin{figure}[htbp]
    \centering
    \begin{subfigure}[b]{0.19\textwidth}
        \includegraphics[width=\textwidth]{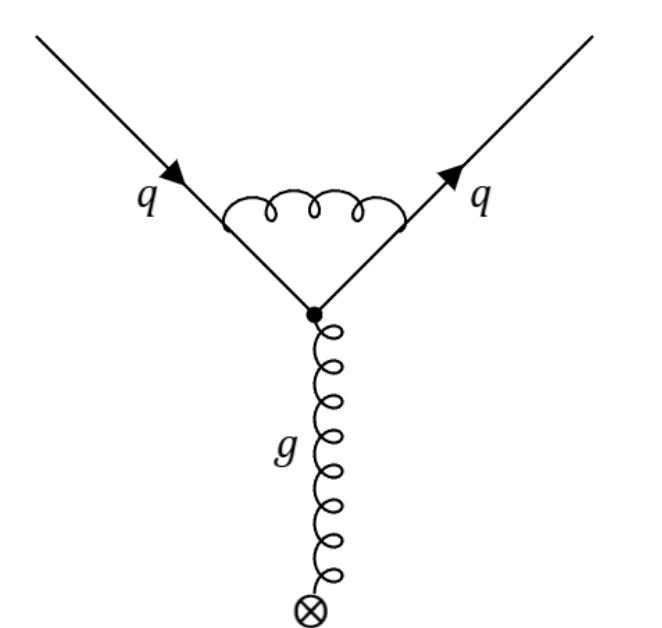}
        \label{fig:img1}
    \end{subfigure}
    \hfill
    \begin{subfigure}[b]{0.19\textwidth}
        \includegraphics[width=\textwidth]{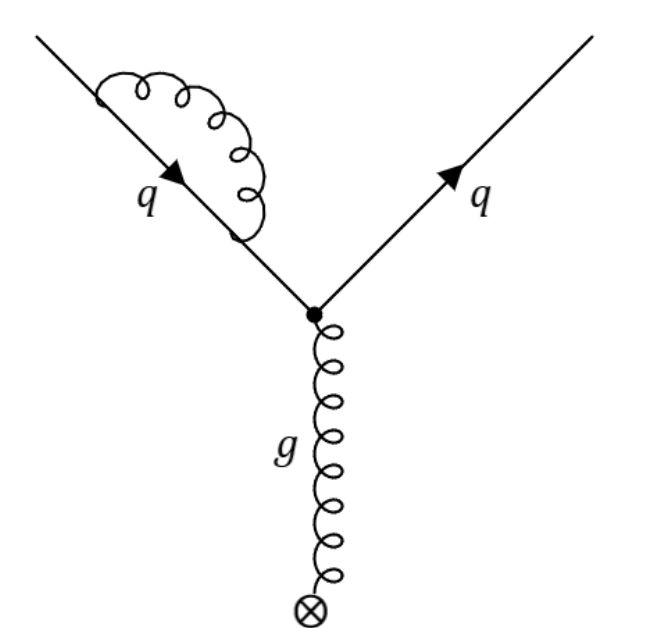}
        \label{fig:img2}
    \end{subfigure}
    \hfill
    \begin{subfigure}[b]{0.19\textwidth}
        \includegraphics[width=\textwidth]{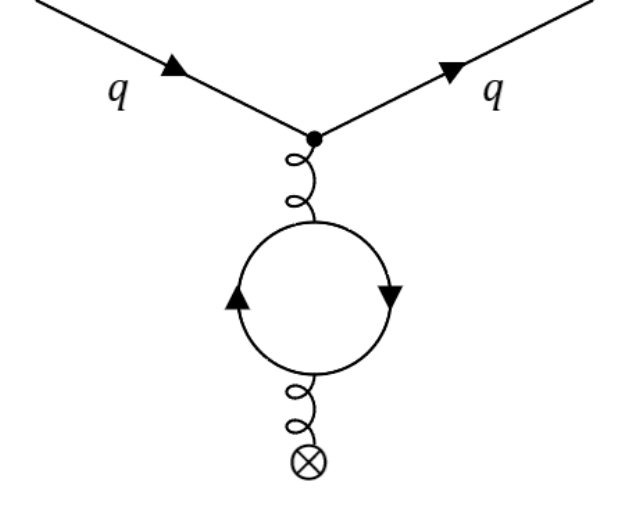}
        \label{fig:img3}
    \end{subfigure}
    \hfill
    \begin{subfigure}[b]{0.19\textwidth}
        \includegraphics[width=\textwidth]{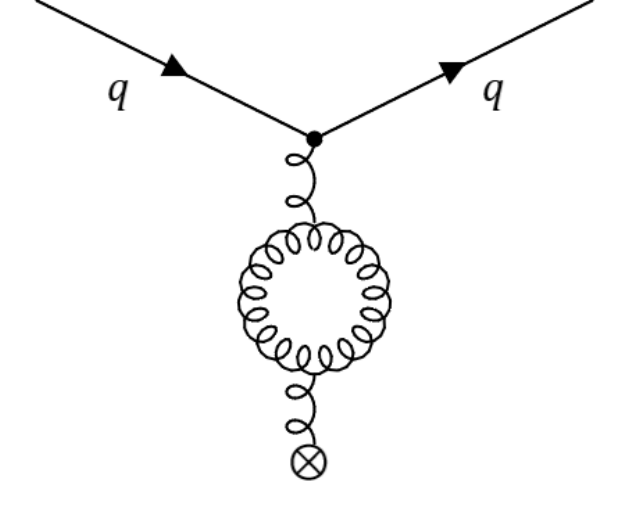}
        \label{fig:img4}
    \end{subfigure}
    \hfill
    \begin{subfigure}[b]{0.19\textwidth}
        \includegraphics[width=\textwidth]{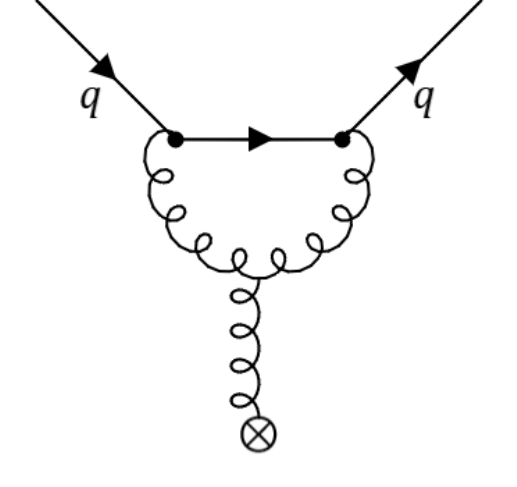}
        \label{fig:img5}
    \end{subfigure}
    \caption{Leading order diagrams contributing to the correction of the heavy quark external potential.}
    \label{fig:feynmanloop}
\end{figure}

\subsection{The Bound State Wave Functions}

In the infinite heavy quark mass limit $M\to \infty $ only the pure coulombic part of the potential survives. The stationary Dirac equation for the light antiquark in the heavy quark potential is:
\begin{equation}
\label{Diraceq}
   \mathcal{H}\psi= (\boldsymbol{\alpha}_{D} \cdot \boldsymbol{p}  + \beta_{D}  m + V(r) )\psi=E \psi \ ,
\end{equation}
where we renamed the light antiquark momentum as $\boldsymbol{p}$.
  The strength of the Coulomb-like potential in Eq. \eqref{Diraceq} carries information about the underlying perturbative IRFP via the value of the coupling constant at the fixed point for $n_f = N_F-2$ massless flavors: 
\begin{equation}
\label{HQ_potential}
    V(r) = -\frac{\alpha}{r}\ , \quad \text{with} \quad \alpha= \frac{4}{3} \alpha_s^*(n_f) \ .
\end{equation}
We can generalize the result to quarks in an arbitrary representation  $r$ of the $SU(N)$ gauge group by replacing $C_F=4/3$ with $C_r$.

We now discuss the solutions of the Dirac equation \eqref{Diraceq}.
Since the angular momentum and parity operators of a charged particle with spin commute with the Hamiltonian, the physical states are characterized by the energy, angular momentum, and parity. The spinor wave functions for stationary states are \cite{Bjorken:1965sts,Greiner:1990tz}
\begin{equation}
    \label{spinor}
    \psi_{njm}= \begin{pmatrix} \varphi_{njlm}  \\ \chi_{njl'm}  \end{pmatrix} \, = \begin{pmatrix} ig_n(r) \Omega_{jlm} \\  f_n(r) \Omega_{jl'm} \end{pmatrix} \ ,
\end{equation}
where $j$ and $l$ are respectively the total and orbital angular momentam of the light antiquark, $m$ corresponds to the projection of the total angular momentum on the quantisation axis, $n$ is the principal quantum number. The orbital angular momentum $l'$ associated with $\chi_{jl'm}$ is defined as \begin{equation}
    l'=2j-l=
\begin{cases}
    l+1 \quad &\text{for} \quad j=l+1/2\ ,\\
    l-1 \quad &\text{for} \quad j=l-1/2\ . 
\end{cases}
\end{equation}
The spherical spinors $\Omega_{jlm}$ are defined, for example, in \cite{Bjorken:1965sts,Greiner:1990tz}. For the useful cases with $j =l\pm 1/2$, we have: 
\begin{equation}
\label{spherical_spinors}
    \begin{aligned}
      & \Omega_{l+\frac{1}{2}lm}=  \begin{pmatrix} \sqrt{\frac{j+m}{2j}} Y_{l m-\frac{1}{2}} \\  \sqrt{\frac{j-m}{2j}} Y_{l m+\frac{1}{2}}\end{pmatrix}\ , \\
        & \Omega_{l-\frac{1}{2}lm}=  \begin{pmatrix} -\sqrt{\frac{j-m+1}{2j+2}} Y_{l m-\frac{1}{2}} \\  \sqrt{\frac{j+m+1}{2j+2}} Y_{l m+\frac{1}{2}}\end{pmatrix}\ , 
    \end{aligned}
\end{equation}
where $Y_{lm}$ are the well-known spherical harmonics. 
In terms of the two-component fields, Eq. \eqref{Diraceq} reads
\begin{equation}
    \label{Dirac_MatrixForm} 
\begin{pmatrix} m+V(r) & \boldsymbol{\sigma} \cdot \boldsymbol{p} \\ \boldsymbol{\sigma} \cdot \boldsymbol{p}& -m +V(r)  \end{pmatrix} \begin{pmatrix} \varphi_{n l j m}  \\ \chi_{n l^{\prime} j m}  \end{pmatrix} = E \begin{pmatrix} \varphi_{n l j m}  \\ \chi_{n l^{\prime} j m}  \end{pmatrix}  \ .
\end{equation}
Upon defining $G_n(r)=rg_n(r)$ and $F_n(r)=rf_n(r)$, Eq.~\eqref{Dirac_MatrixForm} yields the following coupled equation for the radial components 
\begin{equation}
    \begin{aligned}
\label{radial_Dirac}    
   & \frac{dG_n}{dr}+\frac{k}{r}G_n= (E+m-V(r))F_n \ , \\
   -& \frac{dF_n}{dr}+\frac{k}{r}F_n=(E-m-V(r))G_n \ ,
    \end{aligned}
\end{equation}
with 
\begin{equation}
    k=
\begin{cases}
    -(l+1)=-(j+1/2) \quad &\text{for} \quad j=l+1/2\ ,\\
     \qquad \quad\; l=+(j+1/2) \quad &\text{for} \quad j=l-1/2\ .
\end{cases}
\end{equation}
The normalised radial wave functions are \cite{Greiner:1990tz}
\begin{equation}
    \label{radial_wavefunc}
    \begin{aligned}
       \begin{pmatrix}
    g_n(r)\\
    f_n(r)
\end{pmatrix}&= \frac{\pm (2\lambda)^{3/2}}{\Gamma(2\rho+1)} \sqrt{\frac{(m \pm E)\Gamma(2\rho +n'+1)}{4m\frac{(n'+\rho)m}{E}\Big(\frac{(n'+\rho)m}{E}-k\Big)n'!}}(2\lambda r)^{\rho-1}e^{-\lambda r}  \\&\times \Big[\Big(\frac{(n'+\rho)m}{E}-k\Big) {}_1F_1(-n',2\rho+1; 2\lambda r) \mp n'\, {}_1F_1(1-n',2\rho+1; 2\lambda r) \Big] \ ,
    \end{aligned}
\end{equation}
where $n'=n-|k|=n-j-1/2$, $\lambda=\sqrt{m^2-E^2}$, $\rho=\sqrt{k^2-\alpha^2}$, and ${}_1F_1(a,b;x)$ is the confluent hypergeometric functions of first kind. \\
The energy eigenvalues are:
\begin{equation}
\label{energy_eigenvalues}
    E_{nj}=m\Bigg[1+\Bigg(\frac{\alpha}{n-(j+1/2)+\sqrt{(j+1/2)^2-\alpha^2}}\Bigg)^2\,\Bigg]^{-1/2}  \ .
\end{equation}
The radiative correction contributions are of order $\mathcal{O}(\alpha^5)$ and therefore not considered here and to this order Eq. \eqref{energy_eigenvalues} yields:
\begin{equation}
    \label{energy_expanded}
    E_{nj}= m\bigg\{1-\alpha^2\bigg[\frac{1}{2n^2}+\frac{\alpha^2}{2n^3}\bigg(\frac{1}{j+\frac{1}{2}}-\frac{3}{4n}\bigg)\bigg]\bigg\}+\mathcal{O}(\alpha^5)  \ .
\end{equation}
In Figure \ref{fig:zeroenergies}, we show the eigenenergies against $n_f$ for different values of the quantum numbers $n,j$. 
\begin{figure}
    \centering
\includegraphics[width=0.6\textwidth]{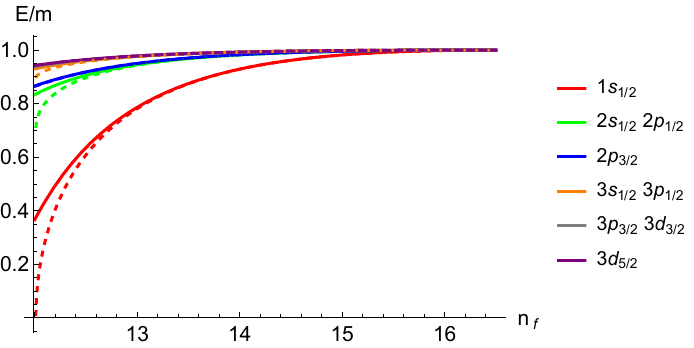}
    \caption{Energy spectrum as function of the number of active massless flavors. Solid and dashed lines refer respectively to Eq. \eqref{energy_expanded} and Eq. \eqref{energy_eigenvalues}. Energy levels are denoted using standard atomic notation (e.g., `s' for $l=0$, `p' for $l=1$, `d' for $l=2$).}
    \label{fig:zeroenergies}
\end{figure}
The normalised spin-up and spin-down wave functions, in the fundamental state $n=1$, $l=0$, $j=1/2$ are given by \cite{Bjorken:1965sts,Greiner:1990tz}:
\begin{equation}
\begin{aligned}
\label{eigenfunctions}
     & \psi_{1,1/2,1/2}= \frac{(2 m \alpha)^{3/2}}{\sqrt{4\pi}} \sqrt{\frac{1+\rho}{2\Gamma(1+2\rho)}}(2m\alpha)^{\rho-1}e^{-m\alpha r}  \begin{pmatrix} 1  \\ 0 \\ \frac{i(1-\rho)}{\alpha}  \cos{\theta} \\ \frac{i(1-\rho)}{\alpha}  \sin{\theta} e^{i \phi } \end{pmatrix}  \ ,\\ 
     & \psi_{1,1/2,-1/2}=\frac{(2 m \alpha)^{3/2}}{\sqrt{4\pi}} \sqrt{\frac{1+\rho}{2\Gamma(1+2\rho)}}(2m\alpha)^{\rho-1}e^{-m\alpha r}  \begin{pmatrix} 0  \\ 1 \\ \frac{i(1-\rho)}{\alpha}  \sin{\theta} e^{-i \phi }  \\ -\frac{i(1-\rho)}{\alpha}  \cos{\theta} \end{pmatrix}  \ ,
\end{aligned}
\end{equation}
where $\rho=\sqrt{1-\alpha^2}$.
The ground-state energy becomes
\begin{equation}
\label{ground_energy}
     E_{1}=m\rho =m\sqrt{1-\alpha^2}  \ .
\end{equation}
In the heavy quark limit of QCD, the difference between the hadron and the
heavy quark mass is given by the parameter $\Bar{\Lambda}$:
\begin{equation}
    \label{Lambdaparameter}
    \Bar{\Lambda}=M_H-M \ ,
\end{equation}
which represents the energy of the brown muck. In our approach, $\Bar{\Lambda}$ can be obtained from the energy eigenvalues of the light antiquark given in Eq. \eqref{energy_eigenvalues}. In particular, for the fundamental state $n=1$, $j=1/2$ we have:
\begin{equation}
    \label{Energy_fundamental}
    \Bar{\Lambda}\equiv E_1=m\sqrt{1-\alpha^2},
\end{equation}
while the binding energy is given by
\begin{equation}
\label{eq:binding_energy}
    E_{bind}=E_1-m=m\bigg(\sqrt{1-\alpha^2}-1\bigg)\approx -\frac{1}{2} m \alpha^2 \ .
\end{equation}
Overall, we are pleased to discover that perturbation theory is able to cover quite a wide range of flavours in the conformal window before higher order corrections are needed as it is clear from figure~\ref{fig:zeroenergies}.  

\subsection{The Conformal Isgur-Wise function}
\label{sec:IWfunctions}
Decays of hadrons containing an heavy quark can be efficiently described in the framework of HQET. For example, let us consider the process $H(v) \to H'(v')$ of hadrons containing an heavy quark. The matrix elements of an
heavy quark current between hadronic states can be expressed as a linear combination of terms, factorized as \cite{Falk:1991nq,Zalewski:1991xb}:
\begin{equation}
    \label{heavy_light_factorize}
     \bra{H'(v')}J^{Q'Q}\ket{H(v)}  = \bra{Q', v', s_{Q'}}J^{Q'Q}\ket{Q, v, s_Q} \bra{ L, J'_{L}}\ket{L, J_{L} }  \ ,
\end{equation}
where $L$ denotes the light degrees of freedom, or brown muck, with angular momentum $J^{(\prime)}_L$, $Q$ and $Q^{\prime}$ denote the heavy quark of the initial and final states. In the infinite mass limit $M\rightarrow \infty$ the transition amplitude associated to the brown muck does not depend on the flavour and spin of the heavy quark and results in a form factor suppression depending only on the velocities $v$ and $v'$ of the heavy quark before and after the transition. The overlap between the states of the brown muck is the Isgur-Wise (IW) function \cite{Isgur:1989vq,Isgur:1990yhj}:
\begin{equation}
    \label{IWfunctioncomp}
    \xi(w)= \sqrt{\frac{2}{w+1}} \bra{L, J'_{L}}\ket{L, J_{L} } \ ,
\end{equation}
where $w=v\cdot v^{\prime}$ is the recoil parameter. \\
The advantage of our theoretical framework is that we can use perturbation theory. Consequently an explicit computation of the IW function becomes, for the first time, feasible. \\
To differentiate our Isgur-Wise function from the non-perturbative one in ordinary QCD, we refer to ours as to the \emph{Conformal Isgur-Wise} (CIW) function.  \\
It is convenient to compute the CIW function in the modified Breit frame \cite{Lie-Svendsen:1987dqs}, where two particles move along the $z$-axis with equal and opposite velocities. The overlap function takes the form 
\begin{equation}
 \label{IWfunctioncomp1}
    \bra{ L, J'_{L}}\ket{L, J_{L} }= \int d^3 x  \,\psi^{\dagger}_{L'}(x) \psi_{L}(x) |_{t =0}\ ,
\end{equation}
where
\begin{equation}
    \psi_{L^{(\prime)}}(x)= \psi_{L^{(\prime)}}(\boldsymbol{x}_\pm)e^{-iE t}  \ ,
\end{equation}
is the wave function of the light degrees of freedom of the initial (final)  hadron, moving in the positive (negative) direction along $z$-axis with velocity $v=\abs{\boldsymbol{v}}$.\\
We use the fact that the wave functions in the initial and final states are related to those in their respective rest frames, denoted as $\psi_{L}^0(x)$, via the Lorentz boost operators $S(\mp\boldsymbol{v})$, corresponding to the Lorentz transformations $\Lambda^{-1}(\mp \boldsymbol{v})\,x_\pm|_{t=0}=(\pm \gamma v z,x,y,\gamma z)$:
\begin{equation}
    \psi_{L^{(\prime)}}(x)=S(\mp\boldsymbol{v})  \psi_{L^{(\prime)}}^{0}(x,y,\gamma z)e^{\mp i E\gamma v z}  \ .
\end{equation}
Plugging back in Eq. \eqref{IWfunctioncomp1} and using the fact that $S^\dagger(\boldsymbol{v})=S(\boldsymbol{v})=S^{-1}(-\boldsymbol{v})$, Lorentz boost operators cancel out and we obtain \cite{Sadzikowski:1993iv,Olsson:1994us,Ahmady:1994ci}
\begin{equation}
 \label{IWfunctioncomp2}
    \bra{L, J'_{L}}\ket{L, J_{L} }= \int d^3 x   \,\psi_{L'}^{0\,\dagger}(x ,y ,\gamma z ) \psi^0_{L}(x ,y ,\gamma z ) e^{-2iE\gamma v z } \ .
\end{equation}
Finally, rescaling the coordinate $z \rightarrow z /\gamma$ and using the kinematic identities 
\begin{equation}
\begin{aligned}
    &\gamma= \sqrt{\frac{w+1}{2}} \ ,\\
   &  v=\sqrt{\frac{w-1}{w+1}} \ ,
\end{aligned}
\end{equation}
 we arrive at the following expression \cite{Sadzikowski:1993iv,Olsson:1994us}
 \begin{equation}
     \xi(w)=\frac{2}{w+1}\int d^3 x  \,\psi_{L}^{0\,\dagger}(x,y,z) \psi^0_{L}(x,y,z) e^{-2iE \sqrt{\frac{w-1}{w+1}}  z} \ .
 \end{equation}
 Using the wave function for the light degrees of freedom given in Eq. \eqref{spinor} and the fact that we are only interested in s-waves ( e.g. $l=0$ ), we obtain the CIW function
 \begin{equation}
 \label{IWfinal}
     \xi(w)= \frac{2}{w+1}\int d r \, r^2 (|g_n(r)|^2+|f_n(r)|^2 )j_0(2E \sqrt{\frac{w-1}{w+1}}  r)\ ,
 \end{equation}
 where we used spherical symmetry and the identity
 \begin{equation}
     e^{-ikz}=\sum_{l=0}^{\infty}(2l+1)(-i)^l j_l(kr) Y_{l0} \ .
 \end{equation}
  Here, $j_l(kr)$ denotes the spherical Bessel functions. \\
 Substituting Eq. \eqref{eigenfunctions} in Eq. \eqref{IWfinal}, we find the CIW function for the fundamental state: 
\begin{equation}
\label{IW1}
    \xi(w)=\frac{1}{w+1}\frac{ \left(1+\Phi^2(w)\right)^{-\rho} }{ \Phi(w)} \frac{\sin \left(2 \rho \arctan\left( \Phi(w) \right)\right)}{\rho}\ , 
\end{equation}
where 
\begin{equation}
   \Phi(w)=\sqrt{\frac{w-1}{w+1}}\frac{\rho}{\alpha} \ , \quad \rho =\sqrt{1-\alpha^2}\ .
\end{equation}
In \cref{fig:three_plots} we show the plots of CIW function in terms of the number of active light flavors $n_f<n_f^{\rm{AF}}$, and the recoil parameter $w$. Here we observe that when $\alpha$ increases and $w$ decreases the CIW function increases towards its maximum. Below we will investigate more carefully the various limits. 

\begin{figure}[h]
  \centering
  \begin{subfigure}[b]{0.6\textwidth}
    \centering
    \includegraphics[width=\textwidth]{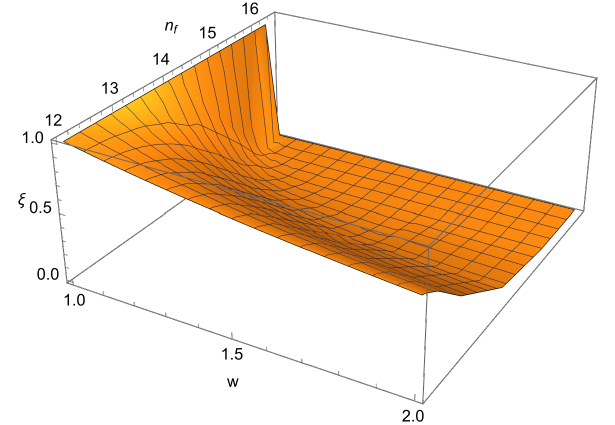}
    \label{fig:3DIW}
  \end{subfigure}
  
  \vspace{1em} 
  
  \begin{subfigure}[b]{0.45\textwidth}
    \centering
    \includegraphics[width=\textwidth]{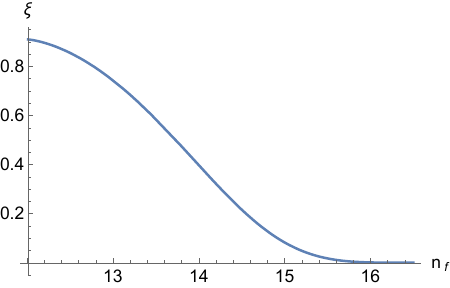}
  \end{subfigure}
  \hfill
  \begin{subfigure}[b]{0.45\textwidth}
    \centering
    \includegraphics[width=\textwidth]{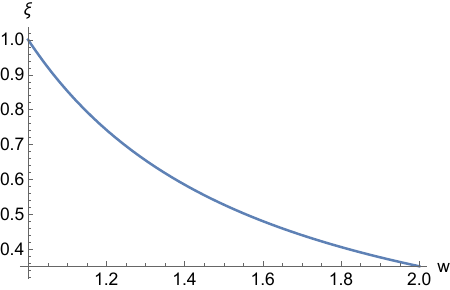}
  \end{subfigure}
  \caption{ CIW function in terms of the recoil parameter $w$ and the number of active massless flavors $n_f$. Left and right bottom panels are obtained fixing $w=1.2$ and $n_f=13$ respectively.}
  \label{fig:three_plots}
\end{figure}
For $\alpha \neq 0$, we recover that the CIW function is normalized to one at zero recoil: $\xi(1)=1$.

Furthermore, by expanding Eq. \eqref{IW1} around the point of zero recoil ($w=1$), we find:
\begin{equation}
\begin{aligned}
    \label{IWexpansion}
  \xi(w) \cong & \,  1+\xi^{\prime}(1)(w-1)+\xi^{\prime \prime}(1)\frac{(w-1)^2}{2} \ , \quad \forall \alpha \neq 0\ ,
\end{aligned}
\end{equation}
with
\begin{equation}
\begin{aligned}
    \label{slope_curvature}
\xi^{\prime}(1)=&-\frac{1}{6 \alpha ^2}\left(-2 \alpha ^4+\left(3 \sqrt{1-\alpha ^2}+2\right) \alpha ^2-3 \left(\sqrt{1-\alpha ^2}+1\right)\right) & \quad \forall \alpha \neq 0 \ , \\
\xi^{\prime\prime}(1)=&\frac{1}{60 \alpha ^4} \Big(4 \alpha ^8-\left(50 \sqrt{1-\alpha ^2}+73\right) \alpha ^2+45 \left(\sqrt{1-\alpha ^2}+1\right)- \\&-\left(20 \sqrt{1-\alpha ^2}+11\right) \alpha ^6+5 \left(5 \sqrt{1-\alpha ^2}+13\right) \alpha ^4\Big) & \quad \forall \alpha \neq 0 \ .
\end{aligned}
\end{equation}
Although the aim of our work is to determine physical properties in the conformal window, it is tempting to compare the slope of the CIW and its curvature parameter with the ones in model computations for QCD performed by several groups \cite{Sadzikowski:1993iv,Olsson:1994us,Ahmady:1994ci,Hogaasen:1994pp,Jenkins:1992se,Das:2016hmv,Bernard:1993ey,ARGUS:1992dng,Blok:1992fc,Mandula:1993sj,PhysRevLett.72.462,PhysRevD.17.3090,Caprini:1997mu,Boyd:1997kz,LeYaouanc:2003rn}. For example, because we work at $\alpha<1$, we obtain the bounds 
\begin{equation}
    \xi^{\prime}(1)<-\frac{1}{2} \ ,
\end{equation}
and  
\begin{equation}
    \xi^{\prime\prime}(1)>\frac{1}{2} \ ,
\end{equation}
suggested via the use of model computations for non-perturbative QCD  \cite{Sadzikowski:1993iv,Olsson:1994us,Ahmady:1994ci}.
The limit where $\alpha$ approaches zero requires separate consideration, as we expect the bound state itself to no longer exist. This scenario, however, can be safely investigated by taking the double limit where $\alpha \rightarrow0$ and $w\rightarrow 1$, while maintaining fixed the ratio:
\begin{equation}
    \lambda=\frac{w-1}{2\alpha^2} \ .
\end{equation}
From Eq. \eqref{IW1}, we find
\begin{equation}
\label{IW3}
    \xi(\lambda)=\frac{1}{(1+\lambda)^2} \ .
\end{equation}
We can now study the two limits: $\lambda \rightarrow 0$ and $\lambda \rightarrow \infty$. \\
The former corresponds to $w$ approaching 1 faster than $\alpha$ approaching 0, yielding:
\begin{equation}
    \xi(\lambda) = 1 - 2\lambda + 3\lambda^2 + \mathcal{O}(\lambda^3).
\end{equation}
In this case, we naturally recover the normalization of the CIW function. The other limit corresponds to $\alpha$ approaching 0 faster than $w$ approaching 1. In this limit the CIW function vanishes because  the bound state no longer exists.
\section{The  confinement potential}
\label{confpotential}
So far we considered the case in which only one of the light quarks had a small mass, and we expect that the theory to remain conformal in the deep IR. We now turn our attention to the case in which all the light quarks are given a common mass  $m \ll \Lambda_{RGI}$  (or  $m \ll M$ for $M < \Lambda_{RGI}$ ). Below this common mass the theory is a pure Yang-Mills with an estimated exponentially suppressed one-loop confining scale  
\begin{equation}
\label{eq:lambda-c-definition}
    \Lambda_c=m \exp[-\frac{2\pi}{\beta_0\,\alpha_s(m)}] \ ,
\end{equation}
with $\alpha_s(m) \simeq \alpha_s^{\ast} (N_F-1) $ and $ \Lambda_c \ll m$.
The strong coupling diverges at $\Lambda_c$ and does not reach the IRFP anymore. The running is summarized in Fig.~\ref{fig:BZ-heavymass-lightmass-running}.

\begin{figure}[h]
    \centering
   \begin{tikzpicture}[x=0.32pt,y=0.32pt,yscale=-1,xscale=1]

\draw [color={rgb, 255:red, 208; green, 2; blue, 27 }  ,draw opacity=1 ][line width=1.5]    (309,150) .. controls (541,145) and (370,292) .. (630,295) ;
\draw  [dash pattern={on 0.84pt off 2.51pt}]  (280,182) .. controls (489.46,179.39) and (438.46,226.39) .. (498.46,268.39) ;
\draw  [dash pattern={on 4.5pt off 4.5pt}]  (135,306) -- (133.33,48.33) ;
\draw [color={rgb, 255:red, 208; green, 2; blue, 27 }  ,draw opacity=1 ][line width=1.5]    (98,46) .. controls (99,104) and (101,146) .. (132,150) ;
\draw  [dash pattern={on 4.5pt off 4.5pt}]  (92,305) -- (90.33,42.33) ;
\draw  [dash pattern={on 4.5pt off 4.5pt}]  (466,309) -- (467.33,41.33) ;
\draw  [dash pattern={on 4.5pt off 4.5pt}]  (503.46,310.39) -- (502.33,42.33) ;
\draw [color={rgb, 255:red, 208; green, 2; blue, 27 }  ,draw opacity=1 ][line width=1.5]    (132,150) -- (309,150) ;
\draw  [dash pattern={on 0.84pt off 2.51pt}]  (72,182) -- (280,182) ;
\draw  (5.33,309.03) -- (652.33,309.03)(70.03,39.33) -- (70.03,339) (645.33,304.03) -- (652.33,309.03) -- (645.33,314.03) (65.03,46.33) -- (70.03,39.33) -- (75.03,46.33)  ;
\draw  [dash pattern={on 0.84pt off 2.51pt}]  (70.33,149.67) -- (132,150) ;

\draw (75,319.4) node [anchor=north west][inner sep=0.75pt]    {$\Lambda _{c}$};
\draw (119,319.4) node [anchor=north west][inner sep=0.75pt]    {$m$};
\draw (408,317.4) node [anchor=north west][inner sep=0.75pt]    {$\Lambda _{RGI}$};
\draw (495,319.4) node [anchor=north west][inner sep=0.75pt]    {$M$};
\draw (634,319.4) node [anchor=north west][inner sep=0.75pt]    {$\mu $};
\draw (18.74,40.4) node [anchor=north west][inner sep=0.75pt]    {$\alpha_s $};
\draw (-60,130.4) node [anchor=north west][inner sep=0.75pt]  [font=\tiny]  {$\alpha _{s}^{*}( N_{F} -1)$};
\draw (-19,167.4) node [anchor=north west][inner sep=0.75pt]  [font=\tiny]  {$\alpha _{s}^{*}( N_{F})$};

\end{tikzpicture}
\begin{tikzpicture}[x=0.32pt,y=0.32pt,yscale=-1,xscale=1]

\draw  [dash pattern={on 4.5pt off 4.5pt}]  (129,354) -- (130,93) ;
\draw [color={rgb, 255:red, 208; green, 2; blue, 27 }  ,draw opacity=1 ][line width=1.5]    (98,97) .. controls (99,155) and (101,197) .. (132,201) ;
\draw  [dash pattern={on 4.5pt off 4.5pt}]  (92,352) -- (93,91) ;
\draw  [dash pattern={on 4.5pt off 4.5pt}]  (283,357) -- (284,96) ;
\draw  [dash pattern={on 4.5pt off 4.5pt}]  (499,356) -- (500,95) ;
\draw [color={rgb, 255:red, 208; green, 2; blue, 27 }  ,draw opacity=1 ][line width=1.5]    (132,201) -- (238,201) ;
\draw  [dash pattern={on 0.84pt off 2.51pt}]  (73,235) -- (292,235) ;
\draw [color={rgb, 255:red, 208; green, 2; blue, 27 }  ,draw opacity=1 ][line width=1.5]    (238,201) .. controls (269,202) and (236,238) .. (292,235) ;
\draw [color={rgb, 255:red, 208; green, 2; blue, 27 }  ,draw opacity=1 ][line width=1.5]    (292,235) -- (396.33,235.67) ;
\draw  [dash pattern={on 0.84pt off 2.51pt}]  (396.33,235.67) .. controls (551.33,260.67) and (478.33,343.67) .. (633.33,349.67) ;
\draw [color={rgb, 255:red, 208; green, 2; blue, 27 }  ,draw opacity=1 ][line width=1.5]    (396.33,235.67) .. controls (544.33,256.67) and (484.33,347.67) .. (633.33,349.67) ;
\draw  [dash pattern={on 0.84pt off 2.51pt}]  (72.33,200.33) -- (132,201) ;
\draw  (7.33,357.1) -- (650.33,357.1)(71.63,79) -- (71.63,388) (643.33,352.1) -- (650.33,357.1) -- (643.33,362.1) (66.63,86) -- (71.63,79) -- (76.63,86)  ;

\draw (73,361.4) node [anchor=north west][inner sep=0.75pt]    {$\Lambda _{c}$};
\draw (115,361.4) node [anchor=north west][inner sep=0.75pt]    {$m$};
\draw (264,359.4) node [anchor=north west][inner sep=0.75pt]    {$M$};
\draw (474,358.4) node [anchor=north west][inner sep=0.75pt]    {$\Lambda _{RGI}$};
\draw (620,364.4) node [anchor=north west][inner sep=0.75pt]    {$\mu $};
\draw (24,83.4) node [anchor=north west][inner sep=0.75pt]    {$\alpha _{s}$};
\draw (-25,222.4) node [anchor=north west][inner sep=0.75pt]  [font=\tiny]  {$\alpha _{s}^{*}( N_{F})$};
\draw (-58,180.4) node [anchor=north west][inner sep=0.75pt]  [font=\tiny]  {$\alpha _{s}^{*}( N_{F} -1)$};

\end{tikzpicture}

    \caption{Below the mass of the light quarks, the pure Yang-Mills dynamics generates a confining scale $\Lambda_c$. The strong coupling (red solid curve) does not reach the IRFP and diverges at $\Lambda_c$.}
    \label{fig:BZ-heavymass-lightmass-running}
\end{figure}
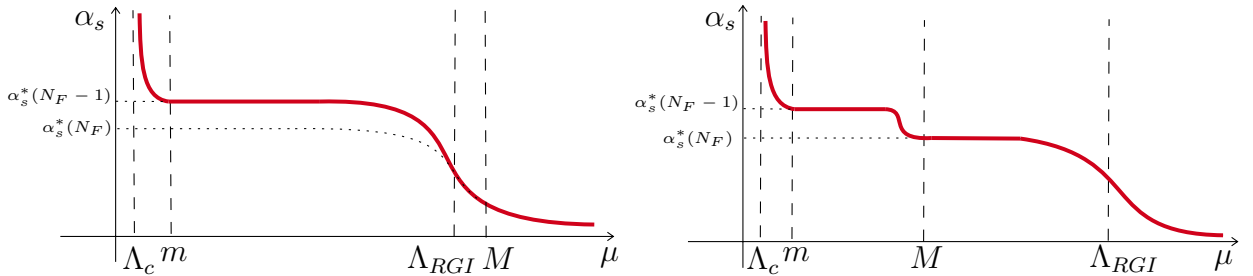
 We can effectively capture the impact of the residual Yang-Mills dynamics  by means of a confinement potential of the form
 \begin{equation}
     V_{conf}(r)=\Lambda_c^2\,r \ .
 \end{equation}
 In general, the confinement term can be introduced in the Dirac equation as the sum of a scalar potential and the time component of a vector potential \cite{Olsson_1995,PhysRevD.17.3090,PhysRevLett.34.369,Matsuki:2004db}. In this analysis, we exclusively focus on the scalar potential to estimate the effects of the confinement term, while avoiding theoretical complexities associated with the Klein paradox induced by a vector-like potential \cite{Olsson_1995}. The modified Dirac equation is:
\begin{equation}
\label{Dirac_string}
   (\boldsymbol{\alpha}_{D} \cdot \boldsymbol{p}  + \beta_{D} m -\frac{\alpha}{r}+\beta_{D} \Lambda_c^2 r)\psi=E \psi \ ,
\end{equation}
where the coupling $\alpha=\frac{4}{3}\alpha^\ast_s$ has to be evaluated at the scale $ m\alpha^2$ of the bound state (see Eq. \eqref{eq:binding_energy}). This introduces  a logarithmic correction to the value of $\alpha^*_s(n_f)$ at $m$ that we neglect. We will see that indeed this is an excellent approximation given the already tiny impact of the confining potential. 

\subsection{Spectroscopy in the presence of Confinement} 
\label{spectroscopy}

The presence of the confinement potential in Eq. \eqref{Dirac_string} affects the binding energies of hadrons.
Being the scale $\Lambda_c$ exponentially suppressed due to the  small coupling constant $\alpha_s(m)\simeq \alpha_s^*(n_f)$, we can study the impact of the confinement potential via ordinary perturbation theory. The corrections to the energy eigenvalues is therefore given by
\begin{equation}
\label{Energies_corr}
     E_{nj}^{(1)}=\bra{\psi^{(0)}_{njm}}\beta_D \Lambda_c^2\,r \ket{\psi^{(0)}_{njm}} =\Lambda_c^2 \int d r \,r^3 (|g_n(r)|^2-|f_n(r)|^2 )\ ,
\end{equation}
where we denoted with $\psi^{(0)}_{njm}$ the unperturbed wave functions associated with the unperturbed energies $E^{(0)}_{nj}$ given in Eq. \eqref{energy_eigenvalues}.
In \cref{fig:corr_energies} we plot the relative variation of the eigenenergies given in Eq. \eqref{Energies_corr} with respect to the unperturbed energies in Eq. \eqref{energy_eigenvalues} as a function of the number of active light flavor $n_f$ for the states with principal quantum number $n=1,2,3$.
\begin{figure}[h]
\centering
\begin{minipage}[b]{0.45\linewidth}
\centering
\includegraphics[width=\textwidth]{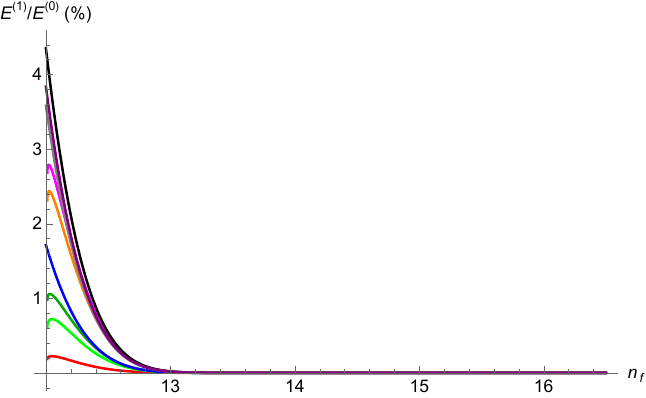}
\end{minipage}
\hfill
\begin{minipage}[b]{0.5\linewidth}
\centering
\includegraphics[width=\textwidth]{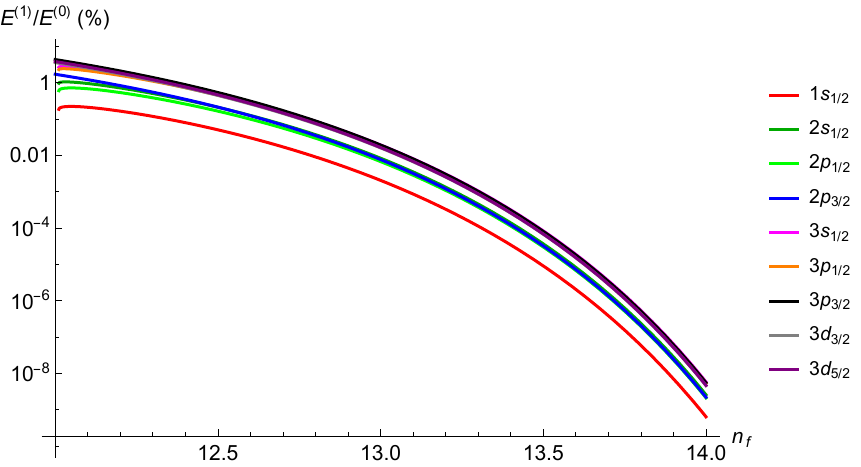}
\end{minipage}
\caption{Relative variation of the energy spectrum compared to the unperturbed result plotted as function of the number of active light flavors. Energy levels are denoted using standard atomic notation (e.g., `s' for $l=0$, `p' for $l=1$, `d' for $l=2$). The right panel shows the logarithmic plot for $12<n_f<14$.}
\label{fig:corr_energies}
\end{figure}

We observe that the larger corrections occur naturally for smaller number of flavours, reaching few percents below 13 flavours. Additionally the light quark wave functions with higher principal number are most affected since they occupy a wider region of space away from the heavy quark where confinement is mostly felt. The strong suppression of the confining potential as we increase the number of light flavours is best appreciated via the right panel of fig.~\ref{fig:corr_energies}. 

\subsection{Confinement corrections to the CIW function}  

The impact of the confinement potential extends to the CIW function and focus  on the ground state to ground state transitions.
The corrected ground-state wave function can be derived using the established formula:
\begin{equation}
\label{confiningenergies}
    \psi_{1,1/2,\pm1/2}= \psi^{(0)}_{1,1/2,\pm1/2}+\sum_{n,j,m\neq 1,1/2,\pm1/2}\frac{\bra{\psi^{(0)}_{njm}}\beta_D \Lambda_c^2 r\ket{\psi^{(0)}_{1,1/2,\pm1/2}}}{E^{(0)}_{1}-E^{(0)}_{nj}}\psi^{(0)}_{njm},
\end{equation}
where the sum runs over all the quantum numbers of the bound states and we neglected the contribution from the continuum spectrum.\\
The first-order correction to the ground state energy $E^{(0)}_1$ is given by
\begin{equation}
\label{eq:corrected_eigenenergies}
     E_1^{(1)}=\Lambda_c^2\frac{2-2\alpha^2+\sqrt{1-\alpha^2}}{2 m \alpha}.
\end{equation}
Then, using orthonormality of spherical harmonics, we observe that only terms with $j=\frac{1}{2}$ contribute to the sum.
Retaining the most relevant order in $\alpha$, we end up with a simple expression in terms of the principal quantum number $n$:

\begin{equation}
\label{eq:perturbative_correction_eigenstates}
  \frac{\bra{\psi^{(0)}_{njm}}\beta_D \Lambda_c^2 r\ket{\psi^{(0)}_{1,1/2,\pm1/2}}}{E^{(0)}_{1}-E^{(0)}_{n}}\approx\frac{16\,\Lambda_c^2}{m^2\alpha^3} \frac{n^{9/2} (n-1)^{n-3}}{(n+1)^{n+3}} \ .
\end{equation}
 It is evident that Eq.  \eqref{eq:perturbative_correction_eigenstates} decreases monotonically with $n$. Therefore we write the first order corrected ground-state wave function, considering only the $n=2$ contribution in the series:
\begin{align}
\label{eq:corrected_eigenfunction}
    \psi_{1,1/2,\pm1/2}=\, \psi^{(0)}_{1,1/2,\pm1/2}+\frac{\Lambda_c^2}{m^2}\,\frac{256 \sqrt{2}}{243\,\alpha^3}\,\psi^{(0)}_{2,1/2,\pm1/2} \ .
\end{align}

The confinement corrected CIW is
\begin{equation}
     \xi(w)=\frac{2}{w+1}\int d^3 x  \,\psi_{1,1/2,\pm1/2}^{\dagger}(x,y,z) \psi_{1,1/2,\pm1/2}(x,y,z) e^{-2i\,E_1 \sqrt{\frac{w-1}{w+1}} z} \ ,
\end{equation}
where the  eigenfunction in Eq. \eqref{eq:corrected_eigenfunction} and eigenenergy in Eq. \eqref{eq:corrected_eigenenergies} appear. To concisely express the modified CIW function, it is useful to define the dimensionless ratio $\Lambda_c / m  = \eta$ and introduce the following functions of the coupling constant $\alpha = \frac{4}{3} \alpha_s^\ast(n_f)$:
\begin{align}\Sigma_0\,=\,&\sqrt{1-\rho} \ , \qquad \qquad  \rho = \sqrt{1-\alpha^2} \notag   \ ,\\
    \Sigma_1\,=\,&\sqrt{\frac{(1-\rho)(\sqrt{2}-\sqrt{1+\rho})\Gamma(1+\rho)\Gamma(2+2\rho)}{(1+\rho)\,(1+\sqrt{2+2\rho})}} \notag \ ,\\
\Sigma_2\,=\,&\sqrt{\frac{(\sqrt{2}+\sqrt{1+\rho})\Gamma(1+\rho)\Gamma(2+2\rho)}{(1+\sqrt{2+2\rho})}} \notag \ ,\\
\Phi(w)\,=\,&\sqrt{\frac{w-1}{w+1}}\frac{\rho}{\alpha} \ . 
\end{align}
We retain terms up to and included order $\mathcal{O}\left( \Lambda_c^2/m^2 \right)$ and obtain

\begin{equation}
    \Delta\xi(w)= \, \xi(w)-\xi^{(0)}(w) =   A(\omega) \, \frac{\Lambda_c^2}{m^2} + {\cal O} \left( \frac{\Lambda_c^4}{m^4} \right) \ ,
\end{equation}
where $\xi^{(0)}(w)$ is the unperturbed CIW given in Eq. \eqref{IW1} and 
\begin{align}
    A(w)= \,& \frac{1}{w+1} \frac{( 2\rho +1) \left(1+\Phi^2(w)\right)^{-\rho } }{\rho\,\alpha\,\Phi(w)}\times\notag \\
    &\left\{\frac{\rho  \Phi(w)}{ 1+\Phi^2(w)}  \cos \left(2 \rho  \arctan \Phi(w) \right)-\left(\frac{\rho  \Phi^2(w) }{ 1+\Phi^2(w)}+\frac{1}{2}\right) \sin \left(2 \rho  \arctan\Phi(w) \right)\right\} + \notag \\
    &\frac{2^{\rho/2}}{(w+1)\Phi(w)}\bigg(\frac{1+\rho}{2}\bigg)^{1/4}\frac{\Gamma(2\rho)}{\Gamma^2(1+2\rho)}\frac{(\Sigma_0\,\alpha)^{-1/2+\rho}}{(\Sigma_0^2-2\alpha^2) }\bigg(\frac{\Sigma_0+\sqrt{2}\alpha}{2}\bigg)^{-2\rho}\notag \\
    &\bigg(1+\frac{16 \alpha^2}{(\sqrt{2}\Sigma_0+2\alpha)^2}\Phi^2(w)\bigg)^{-1/2-\rho}\bigg\{\sqrt{1+\frac{16 \alpha^2}{(\sqrt{2}\Sigma_0+2\alpha)^2}\Phi^2(w)}\times\notag \\
    &[(\Sigma_0^2
-2\alpha^2)(2\Sigma_0+\sqrt{2}\alpha)\Sigma_1+\sqrt{2}\alpha \Sigma_0^2(2\Sigma_0^2-3)\Sigma_2]\sin{\bigg(2\rho\,\arctan{\frac{4\alpha\,\Phi(w)}{\sqrt{2}\Sigma_0+2\alpha}}\bigg)} \notag \\
&+4\rho \Sigma_0(\Sigma_1+\Sigma_2)\sin{\bigg((1+2\rho)\arctan{\frac{4\alpha \Phi(w)}{\sqrt{2}\Sigma_0+2\alpha}}\bigg)}\bigg\}\frac{256\sqrt{2}}{243\alpha^3}  \ .
\end{align}
In \cref{fig:corrIW} we display the impact of the confining potential on the CIW by plotting the relative variation in the CIW function for different values of the number of active flavours $n_f$ and the recoil parameter $w$.
\begin{figure}[h]
  \centering
  \begin{subfigure}[b]{0.7\textwidth}
    \centering
    \includegraphics[width=\textwidth]{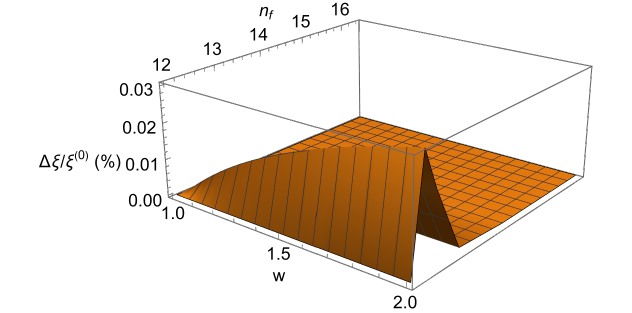}
  \end{subfigure}
  
  \vspace{1em} 
  
  \begin{subfigure}[b]{0.45\textwidth}
    \centering
    \includegraphics[width=\textwidth]{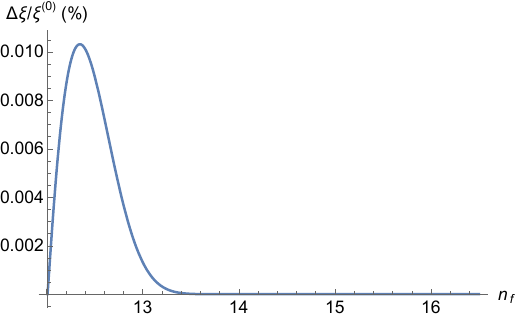}
  \end{subfigure}
  \hfill
  \begin{subfigure}[b]{0.45\textwidth}
    \centering
    \includegraphics[width=\textwidth]{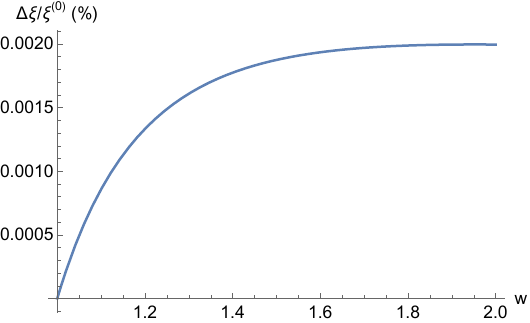}
  \end{subfigure}
  \caption{ Relative variation in the CIW function for different values of the number of active flavours $n_f$ and the recoil parameter $w$. Left and right bottom panels respectively refer  to the sections obtained for $w=1.2$ and $n_f=13$ . }
  \label{fig:corrIW}
\end{figure}
As expected as we approach the upper limit of the conformal window the confining corrections disappear exponentially fast while they start growing around 13 flavours. This is compatible with what we observed at the spectrum level.

\newpage

\section{Conclusions}
\label{conclusions}

We have determined the heavy meson spectrum and dynamics in the perturbative regime of the QCD conformal window at the leading order in the heavy quark mass expansion. As an interesting dynamical quantity, relevant for heavy meson decays, we computed the conformal Isgur-Wise function. We have further determined the impact of the residual low energy confining dynamics stemming from an exponentially suppressed Yang-Mills sector on the heavy meson spectrum.  The results can be straightforwardly generalized to different matter representations. The overall framework  can be readily employed to systematically  account for the dynamics of finite mass defects in generic bulk conformal field theories. Another interesting research avenue is the study of the conformal window  when increasing the charge of the defect. Our findings can be tested via first principle lattice simulations, offering an independent way to acquire new knowledge on the conformal dynamics. Additionally, the results of this work can be of use when constructing dark and bright extensions of the standard model, such as  dark sectors featuring dark heavy meson states. 

\subsection*{Acknowledgments}
F.S. thanks Ofer Aharony, Zohar Komargodski, Matthias Neubert and Robert Pisarski for illuminating discussions and comments on the manuscript. We furthermore thank Giulia Muco for  valuable discussions on the defect conformal field theory properties. F.S. thanks the Institute for Advanced Study at Princeton, the Simons Center for Geometry and Physics and Brookhaven National Laboratory for their generous hospitality while this work was being finalized. The work of F.S. is partially supported by the Carlsberg Foundation, grant CF22-0922.

\printbibliography[heading=bibintoc]
\end{document}